\documentclass[vecphys]{svmult_pgk}		

\usepackage{makeidx}         
\usepackage{graphicx}        
\usepackage{epsfig}          
\usepackage{multicol}        
\usepackage[bottom]{footmisc}
\usepackage{subfigure}
\usepackage{amsmath}
\usepackage{color}
\makeindex             
\begin{document}

\title*{Multiple fluxon analogues and dark solitons in linearly coupled Bose-Einstein condensates}
%
%
\titlerunning{Multiple fluxon analogues and Dark solitons in linearly coupled BECs}
\author{
M.I.\ Qadir$^{1,2}$
\and
H.\ Susanto$^1$
\and
P.C.\ Matthews$^1$
}

\institute{
{$^1$ School of Mathematical Sciences, University of Nottingham,\\
University Park, Nottingham NG7 2RD, UK}\\
{$^2$ Department of Mathematics, University of Engineering and Technology,\\ Lahore, Pakistan}\\
{hadi.susanto@nottingham.ac.uk}\\
{mirfan@uet.edu.pk}
}
\maketitle






Two effectively one-dimensional parallel coupled Bose-Einstein condensates in the presence of external potentials are studied. The system is
modelled by linearly coupled Gross-Pitaevskii equations. In particular, the interactions of grey-soliton-like solutions representing analogues of superconducting Josephson fluxons as well as coupled dark solitons are discussed. A theoretical approximation based on variational formulations to calculate the oscillation frequency of the grey-soliton-like solution is derived and a qualitatively good agreement is obtained.

\section{Introduction}
The concept of electron tunnelling between two superconductors separated by a thin insulating barrier predicted by Josephson \cite{jose62} has been extended relatively recently to tunnelling of Bose-Einstein condensates (BECs) across a potential barrier by Smerzi et al.\ \cite{smer97,ragh99,giov00}. Such tunnelling has been observed experimentally where a single \cite{albi05,levy07} and an array \cite{cata01} of \emph{short} Bose-Josephson junctions (BJJs) were realized. The dynamics of the phase difference between the wavefunctions of the condensates \cite{smer97,ragh99,giov00,ostr00,anan06,jia08} resembles that of point-like Josephson junctions \cite{baro82}.

Recently a proposal for the realization of a \emph{long} BJJ has been presented by Kaurov and Kuklov \cite{kaur05,kaur06}. Similarly to superconducting long Josephson junctions, one may also look for an analogue of Josephson fluxons \cite{usti98} in this case. It was shown in \cite{kaur05,kaur06} that fluxon analogues are given by coupled dark-soliton-like solutions, as the relative phase of the solutions has a kink shape with the topological phase difference equal to $2\pi$. Moreover, it was emphasized that fluxon analogues (FAs) can be spontaneously formed from coupled dark solitons due to the presence of a critical coupling at which the two solitonic structures exchange their stability. The idea of FAs in tunnel-coupled BECs is then extended to rotational FAs in the ground state of rotating annular BECs confined in double-ring traps \cite{bran}. The work in \cite{kaur05,kaur06} was extended in \cite{Qadir12} where the existence and the stability of both FAs and the coupled dark solitons were investigated in the presence of a magnetic trap.

In this chapter, we consider the existence and the stability of multiple FAs and dark solitons in two coupled cigar-shaped condensates in the presence of a magnetic trap along the elongated direction modelled by the normalized coupled Gross-Pitaevskii equations
\begin{equation}
\begin{array}{lll}
\displaystyle i{\psi_j}_t&=&-\frac{1}{2}{\psi_j}_{xx}+|\psi_j|^2\psi_j-\rho_0\psi_j-k\psi_{3-j}+V\psi_j,\\
\end{array}
\label{nls2}
\end{equation}
where $\psi_j,\,j=1,2,$ is the bosonic field, and $t$ and $x$ are the time and axial coordinate, respectively. Here, we assume that the parallel quasi one-dimensional BECs are linked effectively by a weak coupling $k$. Note that herein $k>0$. The case $k<0$ corresponds to an excited state in which there is a $\pi$-phase difference between the condensates. $\rho_0$ is the chemical potential which is considered to be the same in both waveguides and $V$ is the magnetic trap with strength $\Omega$, i.e.\
\begin{equation}
V(x)=\frac{1}{2}\Omega^2 x^2.
\label{ep}
\end{equation}

Different works have been done in similar settings as (\ref{nls2}), e.g. spontaneous symmetry breaking were analyzed in \cite{boris07} when BECs are loaded in two parallel quasi-one-dimensional traps fitted with optical lattices. Before the experimental birth of BEC, the similar settings have been used in the study of stable defects in nonlinear patterns known as optical domain walls \cite{boris94}.  The investigation of the separation of two far separated domain walls along with their stability limits were considered in \cite{boris1994}. Recently, the studies were extended for the case when two components of BECs are coupled by both linear and nonlinear terms \cite{boris11}.

When $\Omega=0$, writing $\psi_j=|\psi_j|\exp(i\varphi_j),$ it was shown that the relative phase $\phi=\pm(\varphi_2-\varphi_1)$ will satisfy a modified sine-Gordon equation \cite{kaur05}. A fluxon analogue of (\ref{nls2}) in that case is given by the solution $\psi_1=\psi_2^*=\psi$, with
\begin{equation}
\psi=\pm \sqrt{\rho_0+k}\tanh(2\sqrt{k}x)\pm i\sqrt{\rho_0-3k}\,\text{sech}(2\sqrt{k}x),
\label{jv}
\end{equation}
where the asterisk denotes complex conjugation. The soliton (\ref{jv}) can be regarded as an analogue of Josephson fluxons \cite{kaur05,kaur06} as the phase difference $\phi$ between the phases of $\psi_1$ and $\psi_2$ forms a spatial kink connecting $\phi=0$ and $\phi=\pm2\pi$. In the following, solution (\ref{jv}) (and its continuations) will be referred to as FAs. From the expression, it is clear that an FA exists only for $0<k<\rho_0/3$. The amplitude of the imaginary part of FA decreases with $k$ and tends to zero as $k \to \rho_0/3$. For $k=\rho_0/3$, the solution in (\ref{jv}) transforms into a dark soliton \cite{kaur05,kaur06}
\begin{equation}
\psi_{1,2}=\pm \sqrt{\rho_0+k}\tanh(\sqrt{\rho_0+k}x),
\label{ds}
\end{equation}
which exists for $k>-\rho_0$. Thus, solutions in (\ref{jv}) and (\ref{ds}) coexist for $0<k<\rho_0/3$. Hence, $k=\rho_0/3$ is a bifurcation point along the family of (\ref{ds}). The bifurcation in this case is a pitchfork bifurcation. When there is no trap, it is found in \cite{kaur05} that the FA solution (\ref{jv}) is stable for all values of $k$ where it exists, while the coupled dark soliton (\ref{ds}) remains unstable for $k<\rho_0/3$ and becomes stable for $k\geq \rho_0/3$.

It is shown in \cite{Qadir12} that the presence of magnetic trap destabilizes the FA solution. However, stabilization is possible by controlling the effective linear coupling $k$ between the condensates. The critical coupling above which FA does not exist is almost independent of the trapping strength. Moreover, the existence and stability regions for coupled dark soliton remain unaffected by the presence of the trap. The transition between FA and dark soliton in the presence of the trap can be realized as a pitchfork bifurcation. In the limiting case, when $\Omega\rightarrow 0$, the critical value of stability $k_{cs}$ of FA goes to zero whereas the critical value of existence $k_{ce}$ remains unaffected.

When the two condensates are uncoupled or the same, i.e.\ $k=0$ or $\psi_1=\psi_2$ with $\rho_0+k \to \rho_0$ respectively, (\ref{nls2}) reduces to
\begin{equation}
\begin{array}{lll}
\displaystyle i \psi_t&=&-\frac{1}{2}\psi_{xx}+|\psi|^2\psi-\rho_0 \psi+V\psi.\\
\end{array}
\label{singlenls}
\end{equation}
In the absence of the external potential, i.e. $V=0$, a single dark soliton on top of a background with constant density $\rho_0$ has the form of \cite{Zakharov73,kivs95} (cf. (\ref{ds}))
\begin{equation}
\psi(x,t)=\sqrt{\rho_0} [A\tanh(\sqrt{\rho_0}A(x-\sqrt{\rho_0} x_0))+iv],
\label{travelds}
\end{equation}
where the parameters $A$ and $v$ determine the soliton depth and velocity respectively and are in general functions of time $t$ with $A^2+v^2=1$. When $v=0$, the dark soliton becomes a stationary kink also called a black soliton and has unit depth (see (\ref{ds})). When $v=1$, the depth of the solution vanishes and the dark soliton becomes the background solution. Since $|\psi|^2$ represents the density of the dark soliton, its minimum value $\rho_{min}$ can be obtained by differentiating $|\psi|^2$ partially with respect to $x$ and equating it to zero, i.e.
\begin{equation}
\frac{\partial |\psi|^2}{\partial x}=0.
\label{1stderivativetest}
\end{equation}
Here,
\begin{equation}
|\psi|^2=\rho_0[A^2\tanh^2(\sqrt{\rho_0}A(x-\sqrt{\rho_0} x_0))+v^2].
\label{density}
\end{equation}
Using Eq. ({\ref{1stderivativetest}}), the critical point we obtain is $x=\sqrt{\rho_0} x_0$. Substituting this value back in Eq. ({\ref{density}}) yields $\rho_{min}=\rho_0 v^2$.

Multiple dark soliton solutions of Eq. (\ref{singlenls}) in the absence of a magnetic trap are also available. The wavefunction for the simplest
case of two dark solitons moving with velocities $v_1=-v_2=v$ can be expressed as \cite{journal2}

\begin{equation}
\psi(x,t)=\frac{(2\rho_0-4\rho_{min})\cosh(qt)-2\sqrt{\rho_0\rho_{min}}\cosh(px)-2iq\sinh(qt)}{2\sqrt{\rho_0}\cosh(qt)+2\sqrt{\rho_{min}}\cosh(px)},
\label{multipleds}
\end{equation}
where $q=2\sqrt{\rho_{min}(\rho_0-\rho_{min})}$ and $p=2\sqrt{\rho_0-\rho_{min}}$.

The dynamics of a dark soliton in BECs in the uncoupled system with magnetic trap has been considered before theoretically \cite{kono08,fran10} (see also \cite{peli05} and references therein) and experimentally \cite{theo10,burg99,beck08,well08,stel08}. Interesting phenomena on the collective behavior of a quantum degenerate bosonic gas, such as soliton oscillations \cite{beck08,well08,theo10} and frequency shifts due to soliton collisions \cite{stel08} were observed. A theoretical analysis based on variational formulation was developed in \cite{kivs95,kivs98} that is in good agreement with numerics as well as with experiments (see, e.g., \cite{fran02,hong09}). A similar variational method was derived in \cite{Qadir12} to explain the dynamics of FAs in (\ref{nls2}). It was shown that the equation of motion for the core of the FA solution is
\begin{eqnarray}
\frac{d^2x_0}{d{t}^2}=\frac{(1-5k)\Omega^2}{1+k}x_0.
\label{x0}
\end{eqnarray}
Note that when $k=1/3$, i.e. the critical coupling for a pitchfork bifurcation between dark solitons and FAs, the oscillation frequency of dark solitons in a harmonic trap is recovered \cite{busc00}(see also \cite{fran02,fran10,theo10}). When the FAs are moving with the velocity $v$, critical value of the coupling constant is \cite{Qadir12}
\begin{equation}
k=-\frac{1}{3}{v}^{2}-\frac{1}{21}+\frac{4}{21}\sqrt{7v^4-7v^2+4}.\label{k_v}
\end{equation}
In a similar fashion as the case when $v=0$, travelling FA solutions are found to be stable in their existing domain. Travelling coupled dark solitons are stable beyond the critical value and unstable otherwise.

Here, we will consider the interaction of multiple FAs as well as dark solitons in (\ref{nls2}) both in the absence and presence of a magnetic trap. Depending on the symmetry of the imaginary parts of the solutions, multiple FAs can be categorized into ($+ -$)-configuration and ($+ +$)-configuration. Note that for dark solitons, we have a single configuration for both ($+ -$) and ($+ +$)-configurations as the imaginary part is zero. In the context of parametrically driven nonlinear Schr\"odinger (NLS) equation, the bound states of FA correspond to Bloch-Bloch states and were discussed in \cite{bara03,bara07}.

The chapter is outlined as follows. In Section 2, we will derive a variational formulation for the oscillation frequency of the ($+ -$)-configuration of FA solution. In Section 3, we will consider the interactions of FAs and dark solitons in (\ref{nls2}) in the absence and presence of a trap. We conclude the work in Section 4.

\section{Variational approximations}
In this section, we shall first derive the interaction potential of two dark solitons given by (\ref{singlenls}) in the absence of the magnetic trap, which was discussed rather briefly in \cite{theo10,kivs95}. We shall then generalize the concept for the interaction of $n$ solitons. The interaction potential will then be used to approximate the oscillation frequency of multiple FAs in the presence of the trap.

\subsection{Determining the interaction potential}
When $k=0$, the system (\ref{nls2}) is decoupled and we are left with the one-dimensional NLS equation (\ref{singlenls}). We consider the interaction of two dark solitons in (\ref{singlenls}) where one of the solitons is located at $x=x_0$ while the other is at $x=-x_0$. Both solitons are moving with velocities equal in magnitude but opposite in signs, i.e. $v_1=-v_2=v$. Then in the weak interacting limit and in the absence of external potential, one can find the equation of the trajectory of the dip of the soliton $x_0$ as a function of time $t$. To do this, we identify the soliton dip $x_0$ as the point of minimum density (cf. Eq. (\ref{1stderivativetest})). In this case, $|\psi|^2$ can be obtained from  Eq. (\ref{multipleds}) and is given by
\begin{equation}
|\psi|^2=\frac{[(2\rho_0-4\rho_{min})\cosh(qt)-2\sqrt{\rho_0\rho_{min}}\cosh(px)]^2+4q^2\sinh^2(qt)}{[2\sqrt{\rho_0}\cosh(qt)+2\sqrt{\rho_{min}}\cosh(px)]^2}. \label{density1}
\end{equation}
Differentiating $|\psi|^2$ partially with respect to $x$ and equating the resulting equation to zero yields
\begin{eqnarray*}
(8\rho^2_0-8\rho^2_0 v^2+16\rho^2_0 v^4-16\rho^2_0 v^2)\cosh^2(qt)+4q^2\sinh^2(qt)&&\\
+(8\rho^2_0 v^3-8\rho^2_0 v)\cosh(px)\cosh(qt)&=&0.
\end{eqnarray*}
Using $A^2=1-v^2$, the above equation can be written as
\begin{eqnarray*}
(8\rho^2_0 A^2-16\rho^2_0 v^2 A^2)\cosh^2(qt)+4q^2\sinh^2(qt)-8\rho^2_0 v A^2\cosh(px)\cosh(qt)=0.
\end{eqnarray*}
Since $q=2\sqrt{\rho_{min}(\rho_0-\rho_{min})}$ or $q^2=4\rho^2_0 v^2 A^2$, substituting the value of $q^2$ in the above equation and using the identity $\cosh^2(qt)-\sinh^2(qt)=1$, we obtain
\begin{eqnarray*}
8\rho^2_0 A^2 \cosh^2(qt)-16\rho^2_0 v^2 A^2-8\rho^2_0 v A^2\cosh(px)\cosh(qt)=0.
\end{eqnarray*}
Dividing throughout by $8\rho^2_0 v A^2 \cosh(qt)$, the equation simplifies to
\begin{eqnarray}
x=\frac{1}{p}\cosh^{-1}\biggr(\frac{\cosh(qt)}{v}-\frac{2v}{\cosh(qt)}\biggr).
\label{dipeq1}
\end{eqnarray}
Then the minimum distance $2x^\ast_0$ between the two dark solitons corresponding to $t=0$ can be obtained from the last equation as
\begin{eqnarray}
2x^\ast_0=\frac{2}{p}\cosh^{-1}\biggr(\frac{1}{v}-2v\biggr).
\label{dipeq2}
\end{eqnarray}

When the solitons are moving slowly, they remain well separated for every value of time. This suggests that the distance $2x^\ast_0$ should be large. This can be justified if the second term on the right hand side of Eq. (\ref{dipeq1}) is much smaller than the first term and hence can be neglected. Then, the resulting equation at $x=x_0$ can be written as
\begin{eqnarray}
x_0=\frac{1}{p}\cosh^{-1}\biggr(\frac{\cosh(qt)}{v}\biggr).
\label{dipeq3}
\end{eqnarray}
Note that differentiating Eq. (\ref{dipeq3}) twice with respect to time yields
\begin{eqnarray*}
\frac{d^2x_0}{dt^2}&=&\frac{A^2 q^2 v^{-3}\cosh(qt)}{p[v^{-2}\cosh^2(qt)-1]^{3/2}}.
\end{eqnarray*}
From Eq. (\ref{dipeq3}), we have $v^{-1}\cosh(qt)=\cosh(px_0)$. Substituting this value and using the identity $\cosh^2(px_0)-1=\sinh^2(px_0)$ in the second derivative above yields
\begin{equation}
\frac{d^2x_0}{dt^2}=\frac{A^2 q^2\cosh(px_0)}{pv^2\sinh^3(px_0)}=-\frac{\partial}{\partial x_0}\biggr(\frac{A^2 q^2}{2 p^2 v^2\sinh^2(px_0)}\biggr).
\label{eqofmotion}
\end{equation}
Eq. (\ref{eqofmotion}) is the equation of motion of the dip of soliton from which we acquire the repulsive potential $W$ to be
\begin{eqnarray*}
W&=&\frac{A^2 q^2}{2 p^2 v^2\sinh^2(px_0)}.
\end{eqnarray*}
Substituting the values of $p$ and $q$, we obtain
\begin{eqnarray}
W&=&\frac{\rho_0 A^2}{2\sinh^2(2\sqrt{\rho_0}A x_0)}.
\label{effectivepotential}
\end{eqnarray}
It is clear that this potential is velocity dependent as $A^2=1-v^2$. Even though the potential $W$ is relevant to the symmetric interactions, however it can be applied to the asymmetric interactions as well, provided that the average depth of the two solitons is used.

The effective repulsive potential (\ref{effectivepotential}) can be used to construct an approximate potential for the interaction of $n$ solitons. In this case the position of the dip of the $i$-th soliton (where $i=1,2,...,n$) is at $x_i$ and is moving with velocity $v_i$ and having depth $A_i=\sqrt{1-v^2_i}$. We may define respectively the average depth and the relative position of the dip for the $i$-th and $j$-th solitons as $A_{ij}= (1/2)(A_i+A_j)$ and $x_{ij}= (1/2)(x_i-x_j)$. Then the repulsive potential $W_i$ can be expressed as
\begin{eqnarray}
W_i&=&\sum_{i\neq j}^n\frac{\rho_0 A_{ij}^2}{2\sinh^2[\sqrt{\rho_0}A_{ij}(x_i-x_j)]}.
\label{effectivepotential1}
\end{eqnarray}

The kinetic energy $E$ and the potential energy $V$ of a structure of $n$ interacting solitons are given by $E=\sum_{i=1}^n (1/2)\dot{x}^2_i$ and $V=\sum_{i=1}^n W_i$. Here dot represents the derivative with respect to time $t$. The Lagrangian $\mathcal{L}$, which is the difference of kinetic and potential energies, is $\mathcal{L}=E-V$. To find the equations of motion we use the Euler-Lagrange equations
\begin{eqnarray}
\frac{\partial}{\partial t} \biggr(\frac{\partial \mathcal{L}}{\partial \dot{x}_i} \biggr)-\frac{\partial \mathcal{L}}{\partial {x_i}}=0, i=1,2,...,n.
\label{EulerLageqs}
\end{eqnarray}
Hence the following $n$ coupled dynamical equations for the trajectories $x_i(t)$ of $n$ interacting solitons are obtained as
\begin{eqnarray}
\ddot{x}_i-\sum_{k=1}^n \biggr(\frac{\partial^2 V}{\partial x_k \partial \dot{x}_i}\dot{x}_k+\frac{\partial^2 V}{\partial \dot{x}_k \partial \dot{x}_i}\ddot{x}_k \biggr)+\frac{\partial V}{\partial x_i}=0.
\label{nEulerLageqs}
\end{eqnarray}

\subsection{Variational approximation for multiple FAs}
We can now use a Lagrangian approach to find the oscillation frequency $\Omega$ of multiple FA solution, in the presence of a magnetic trap. Here, we assume that the FAs are well separated. Then in the limiting case when $k$ is close to the critical coupling for a pitchfork bifurcation, the Lagrangian can be written as
\begin{eqnarray}
\mathcal{L}=\frac{1}{2}(\dot{x}_1^2+\dot{x}_2^2)+\biggr(\frac{1-5k}{1+k}\biggr)\Omega^2 ({x_1^2+x_2^2}) -\frac{ \rho_0}{\sinh^2[\sqrt{\rho_0}(x_2-x_1)]}.
\label{lagr1}
\end{eqnarray}
Note that we have used (\ref{x0}) to describe the potential due to a magnetic trap to an FA.
Since we assume that the FAs are well separated, i.e. $|x_2-x_1|\gg0$, this implies that $e^{-\sqrt{\rho_0} (x_2-x_1)}$  approaches zero. Hence (\ref{lagr1}) can be approximated by
\begin{eqnarray*}
\mathcal{L}=\frac{1}{2}(\dot{x}_1^2+\dot{x}_2^2)+\biggr(\frac{1-5k}{1+k}\biggr)\Omega^2 ({x_1^2+x_2^2})-4 \rho_0 e^{-2\sqrt{\rho_0}(x_2-x_1)}.
\end{eqnarray*}
Using Eq. (\ref{EulerLageqs}), we then have the following system of governing equations
%
\begin{eqnarray}
\ddot{x}_1=-8 \rho_0^{\frac{3}{2}} e^{-2\sqrt{\rho_0}(x_2-x_1)}+2 \biggr(\frac{1-5k}{1+k} \biggr)\Omega^2 x_1,
\label{lagr2}
\end{eqnarray}

\begin{eqnarray}
\ddot{x}_2=8 \rho_0^{\frac{3}{2}} e^{-2\sqrt{\rho_0}(x_2-x_1)}+2 \biggr(\frac{1-5k}{1+k} \biggr)\Omega^2 x_2.
\label{lagr3}
\end{eqnarray}

In order to find the fixed points of this system, we set $\ddot{x}_1=0=\ddot{x}_2$. On adding the resulting equations one can easily see that both fixed points $x_1$ and $x_2$ are additive inverse of each other i.e. $x_1=-x_2=\tilde x$ (say), from which we obtain a single nonlinear algebraic equation which is
\begin{eqnarray}
8 \rho_0^{\frac{3}{2}} e^{4\sqrt{\rho_0}\tilde x}-2 \biggr(\frac{1-5k}{1+k} \biggr)\Omega^2 \tilde x=0.
\label{lagr4}
\end{eqnarray}
We solve this equation numerically to find the values of $\tilde x$ corresponding to different values of $k$.

Now let $\delta_1$ and $\delta_2$ be small perturbations in $x_1$ and $x_2$ and $X_1=x_1+\delta_1(x_1,t)$, $X_2=x_2+\delta_2(x_2,t)$ be the solutions of Eq.(\ref{lagr2}) and Eq.(\ref{lagr3}), respectively. Substituting these solutions into Eq.(\ref{lagr2}) with $X_1=-X_2$, we obtain
\begin{eqnarray*}
\ddot{\delta}_1=-8 \rho_0^{\frac{3}{2}} e^{4\sqrt{\rho_0}\tilde x} e^{-2\sqrt{\rho_0}(\delta_2-\delta_1)}+2 \biggr(\frac{1-5k}{1+k} \biggr)\Omega^2(\tilde x+\delta_1).
\end{eqnarray*}
Using Taylor series expansion of $e^{-2\sqrt{\rho_0}(\delta_2-\delta_1)}$ in the first term on the right hand side yields
\begin{eqnarray*}
\ddot{\delta}_1=-8 \rho_0^{\frac{3}{2}} e^{4\sqrt{\rho_0}\tilde x} [1-2\sqrt{\rho_0}(\delta_2-\delta_1)]+2 \biggr(\frac{1-5k}{1+k} \biggr)\Omega^2(\tilde x+\delta_1).
\end{eqnarray*}
Since $\tilde x$ is a fixed solution, the terms $-8 \rho_0^{\frac{3}{2}} e^{4\sqrt{\rho_0}\tilde x}$ and $2 (\frac{1-5k}{1+k})\Omega^2 \tilde x$ provide only the vertical shift in the solution $\delta_1$, but do not affect the oscillation frequency and hence can be neglected. So, we have
\begin{eqnarray}
\ddot{\delta}_1=16 \rho_0^2 e^{4\sqrt{\rho_0}\tilde x} (\delta_2-\delta_1)+2 \biggr(\frac{1-5k}{1+k} \biggr)\Omega^2 \delta_1.
\label{lagr5}
\end{eqnarray}
Similarly from Eq.(\ref{lagr3}) we obtain
\begin{eqnarray}
\ddot{\delta}_2=-16 \rho_0^2 e^{4\sqrt{\rho_0}\tilde x} (\delta_2-\delta_1)+2 \biggr(\frac{1-5k}{1+k} \biggr)\Omega^2 \delta_2.
\label{lagr6}
\end{eqnarray}

Now let $\omega$ is the common oscillation frequency of FA solutions, then we can write $\delta_1=\gamma_1 e^{i\omega t}$ and $\delta_2=\gamma_2 e^{i\omega t}$. Substituting these values into Eq. (\ref{lagr5}) and Eq. (\ref{lagr6}), we obtain

\begin{eqnarray}
-\omega^2 \gamma_1=16 \rho_0^2 e^{4\sqrt{\rho_0}\tilde x} (\gamma_2-\gamma_1)+2 \biggr(\frac{1-5k}{1+k} \biggr)\Omega^2 \gamma_1,\label{lagr7}\\
-\omega^2 \gamma_2=-16 \rho_0^2 e^{4\sqrt{\rho_0}\tilde x} (\gamma_2-\gamma_1)+2 \biggr(\frac{1-5k}{1+k} \biggr)\Omega^2 \gamma_2.\label{lagr8}
\end{eqnarray}
This system of equations represents an eigenvalue problem and can be written in matrix form as $AY=\lambda Y$, where
\begin{equation*}
A=
\begin{bmatrix}
2(\frac{1-5k}{1+k})\Omega^2-16 \rho_0^2 e^{4\sqrt{\rho_0}\tilde x}  & 16 \rho_0^2 e^{4\sqrt{\rho_0}\tilde x} \\
16 \rho_0^2 e^{4\sqrt{\rho_0}\tilde x}  & 2(\frac{1-5k}{1+k})\Omega^2-16 \rho_0^2 e^{4\sqrt{\rho_0}\tilde x}
\end{bmatrix},
\end{equation*}
$Y=[\gamma_1, \gamma_2]^T$ (T represents the transpose) and $\lambda=-\omega^2$. The characteristic frequency which corresponds to in-phase oscillations (i.e. $\delta_1=\delta_2$) of FA solutions is
\begin{equation}
\omega=\sqrt{2\biggr(\frac{5k-1}{k+1}\biggr)}\Omega,\label{omega1}
\end{equation}
while the frequency corresponding to out-of-phase oscillations (i.e. $\delta_1=-\delta_2$) is
\begin{equation}
\omega=\sqrt{2\biggr(\frac{5k-1}{k+1}\biggr)\Omega^2+32 \rho_0^2 e^{4\sqrt{\rho_0}\tilde x}}.\label{omega2}
\end{equation}
It is important to note that in the above calculations we did not distinguish between ($+ -$) and ($+ +$)-configurations of FAs. It is because the imaginary part of the solution is treated as a passive component. Later through comparisons with numerical results we will see that the theoretical results above are only valid for the ($+ -$)-configuration.
\section{Numerical simulations and computations}
\subsection{Interactions of uncoupled dark solitons without trap}
Let us reconsider the interaction of two dark solitons in (\ref{singlenls}), i.e. (\ref{nls2}) with $k=0$, which are located at $x=\pm x_0$ and are moving with velocities $v_1=-v_2=v$. This problem has been considered in details in \cite{theo10}. Since the domain of inverse hyperbolic cosine is $[1,\infty[$, Eq. (\ref{dipeq2}) holds for $1/v-2v>1$ or $v^2<1/4$, otherwise it gives a complex value for $x_0$. This means that there exists a critical value of velocity $v_{cr}=1/2$ which separates two scenarios.

\begin{figure*}[tbhp!]
\begin{center}
\subfigure[]{\includegraphics[width=8cm]{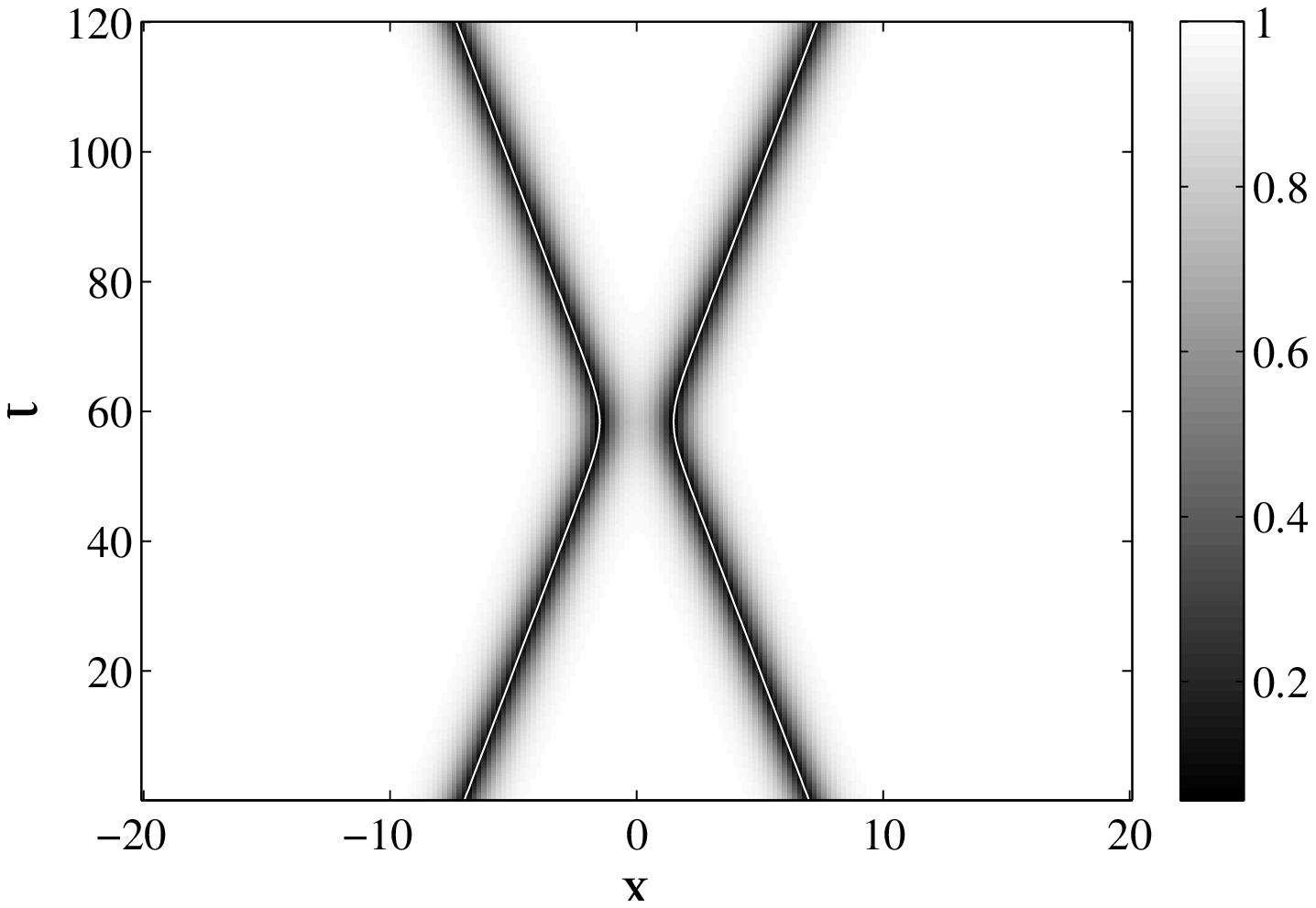}\label{Fig.59a}}
\subfigure[]{\includegraphics[width=8cm]{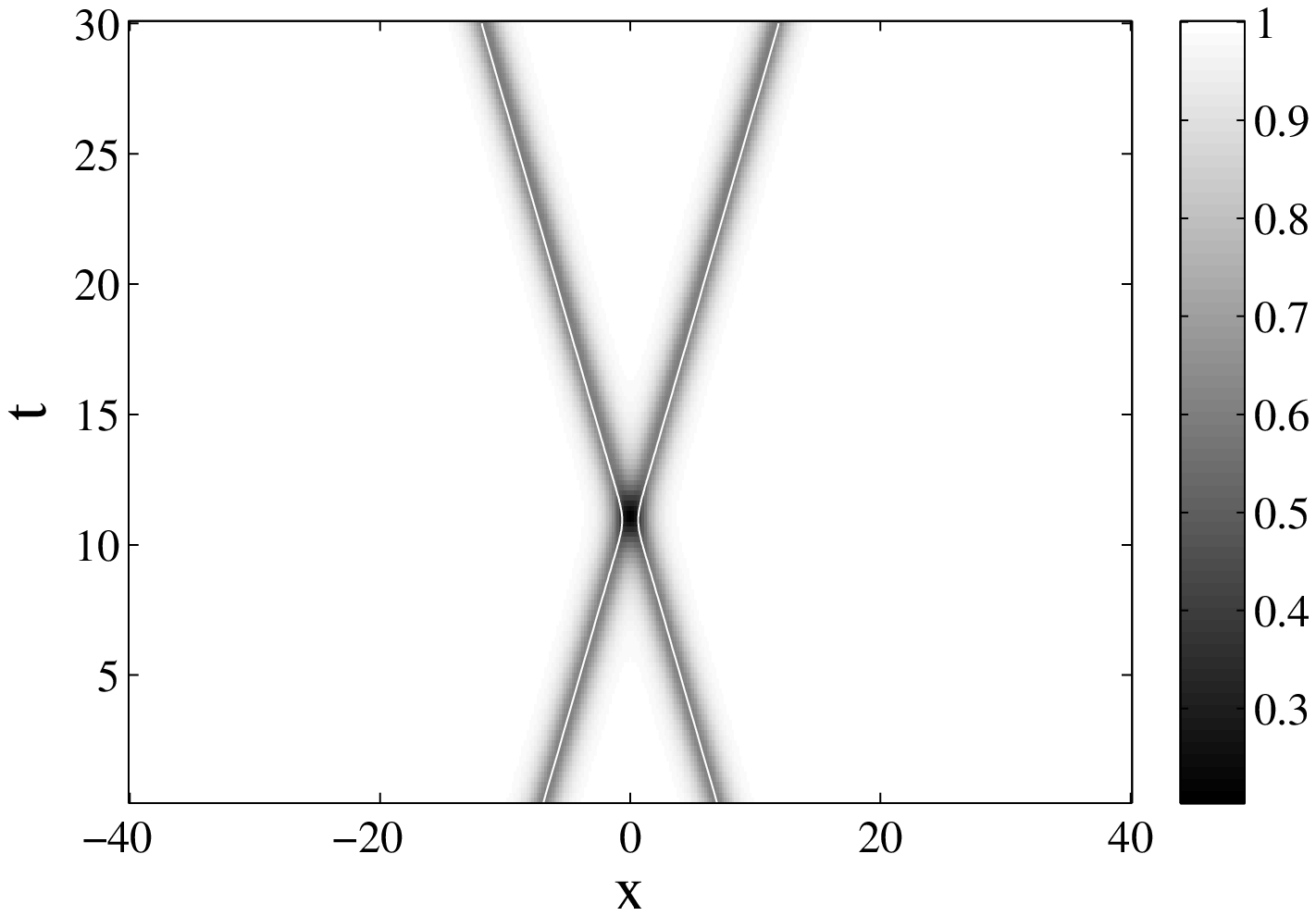}\label{Fig.60a}}
\subfigure[]{\includegraphics[width=8cm]{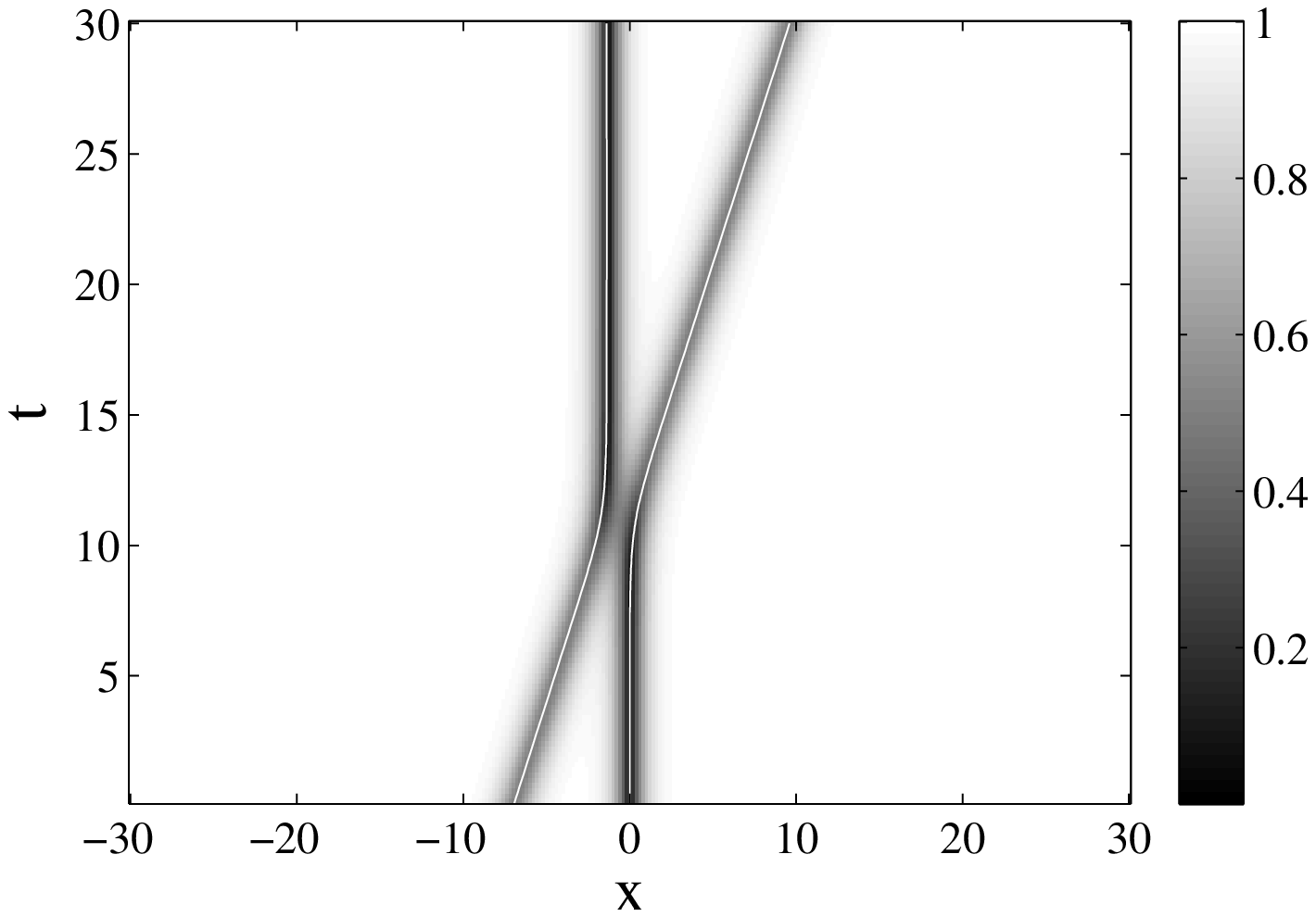}\label{Fig.61a}}
\caption{Numerical evolutions of interaction of two dark solitons. (a) The solitons are moving with velocities $v_1=-v_2=0.1$. The interparticle
repulsion is dominant over the kinetic energies of solitons and solitons are going away from each other after interaction. (b) The solitons are moving with velocities $v_1=-v_2=0.6$. The interparticle repulsion is suppressed by the kinetic energies of the solitons and they transmit through each other at the interacting point. (c) One of the solitons is static and the other is moving with velocity $v=0.5$. After interaction, the travelling soliton becomes stationary and the static soliton starts moving with the velocity of the other soliton. In all panels, the white solid curves are simulations of trajectories of solutions obtained through Eq. (\ref{nEulerLageqs}). Note that solutions in upper and middle panels have an exact analytic solution given by (\ref{multipleds}).}
\label {Fig.59}
\end{center}
\end{figure*}

In the first scenario, two dark solitons having velocities $v_1=-v_2=v<v_{cr}$ start coming close to each other and at the point of their closest proximity, they repel and continuously go away from each other. In this case, before and after the interaction, both dark solitons can be described by two individual density minimum equal to zero.  This shows that dark solitons moving with velocity $v<v_{cr}$ are well separated and can be regarded as low speed solitons. Physically this means that well-separated low speed solitons are repelled by each other and their low kinetic energy could not overcome the interparticle repulsion. A direct numerical integration of (\ref{singlenls}) is performed and shown in Fig.\ \ref{Fig.59a}. Numerical simulations have also been done to check the validity of Eq. (\ref{nEulerLageqs}). The trajectories obtained through Eq. (\ref{nEulerLageqs}) are then plotted in  Fig.\ \ref{Fig.59a} and indicated by white solid curves. The approximation shows excellent agreement qualitatively as well as quantitatively with the results obtained through direct numerical integration of Eq. (\ref{singlenls}).

In the second scenario, dark solitons approaching each other with velocity greater than the critical velocity will collide and after collision transmit through each other. Unlike low speed solitons, at the collision point they overlap entirely and are indistinguishable. Physically this means that due to the high velocity, their kinetic energy defeats the interparticle repulsion. This situation is shown in Fig.\ \ref{Fig.60a}. Even though this case is beyond the particle-like approximation, we show in Fig.\ \ref{Fig.60a} that Eq. (\ref{nEulerLageqs}) can still provide an approximate trajectory of the soliton collision.

In the above discussion, we only considered the symmetric case where solitons collide with the same absolute velocity. Let us now consider the asymmetric case. In this case, a dark soliton moving with velocity $v$ will interact with a static dark soliton. At the interaction point, the static soliton is repelled by the travelling soliton. The energy possessed by the moving soliton is used to push the static soliton away from the original position. The travelling soliton transfers all its kinetic energy to the static soliton and becomes stationary after collision as shown in Fig.\ \ref{Fig.61a}. The white solid lines in this figure depict the trajectories obtained through the numerical integration of Eq. (\ref{nEulerLageqs}).

\subsection{Interaction of dark solitons in coupled NLS equations without trap}
\begin{figure*}[tbhp!]
\begin{center}
\subfigure[]{\includegraphics[width=8cm]{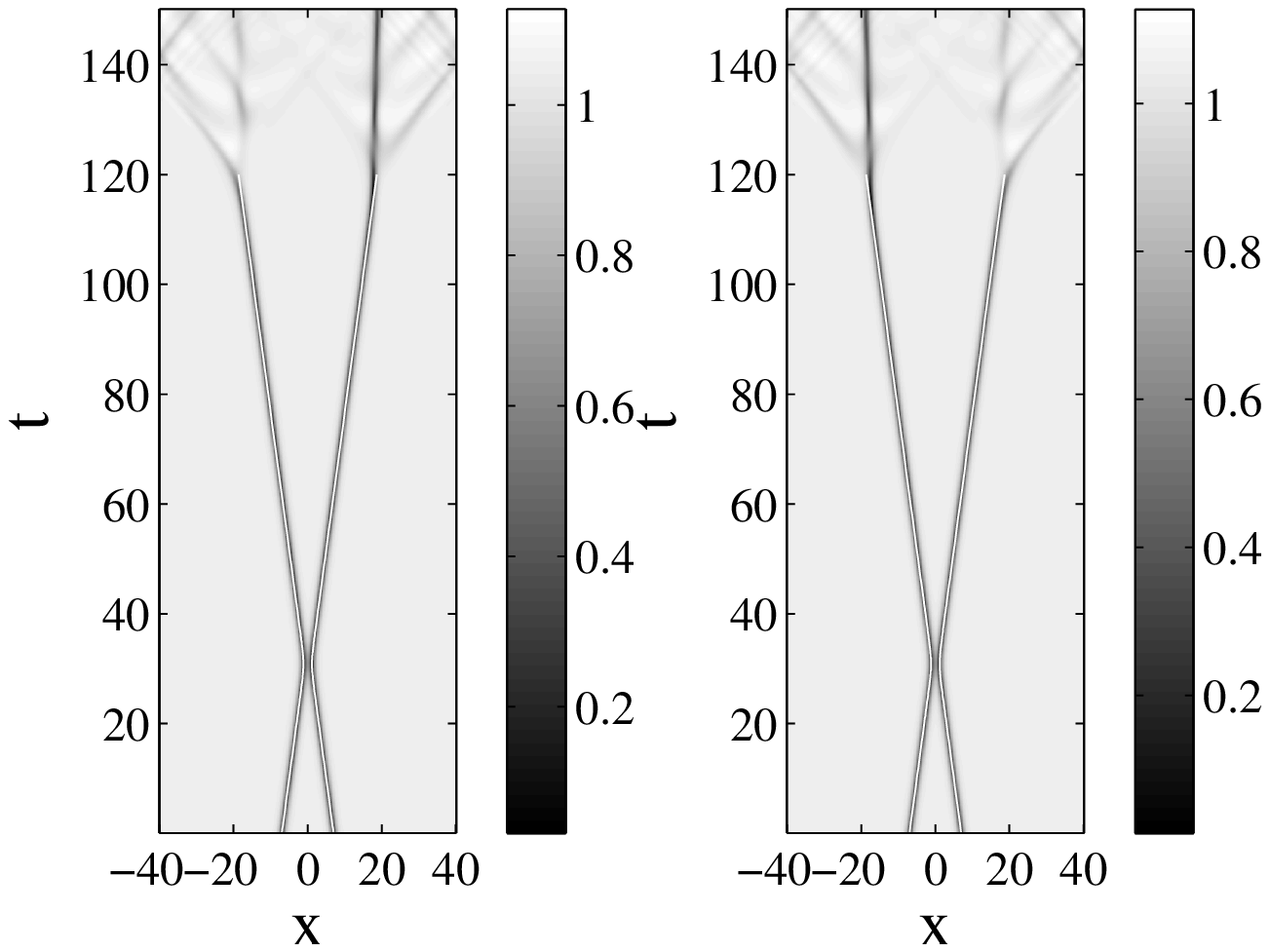}\label{Fig.63a}}
\subfigure[]{\includegraphics[width=8cm]{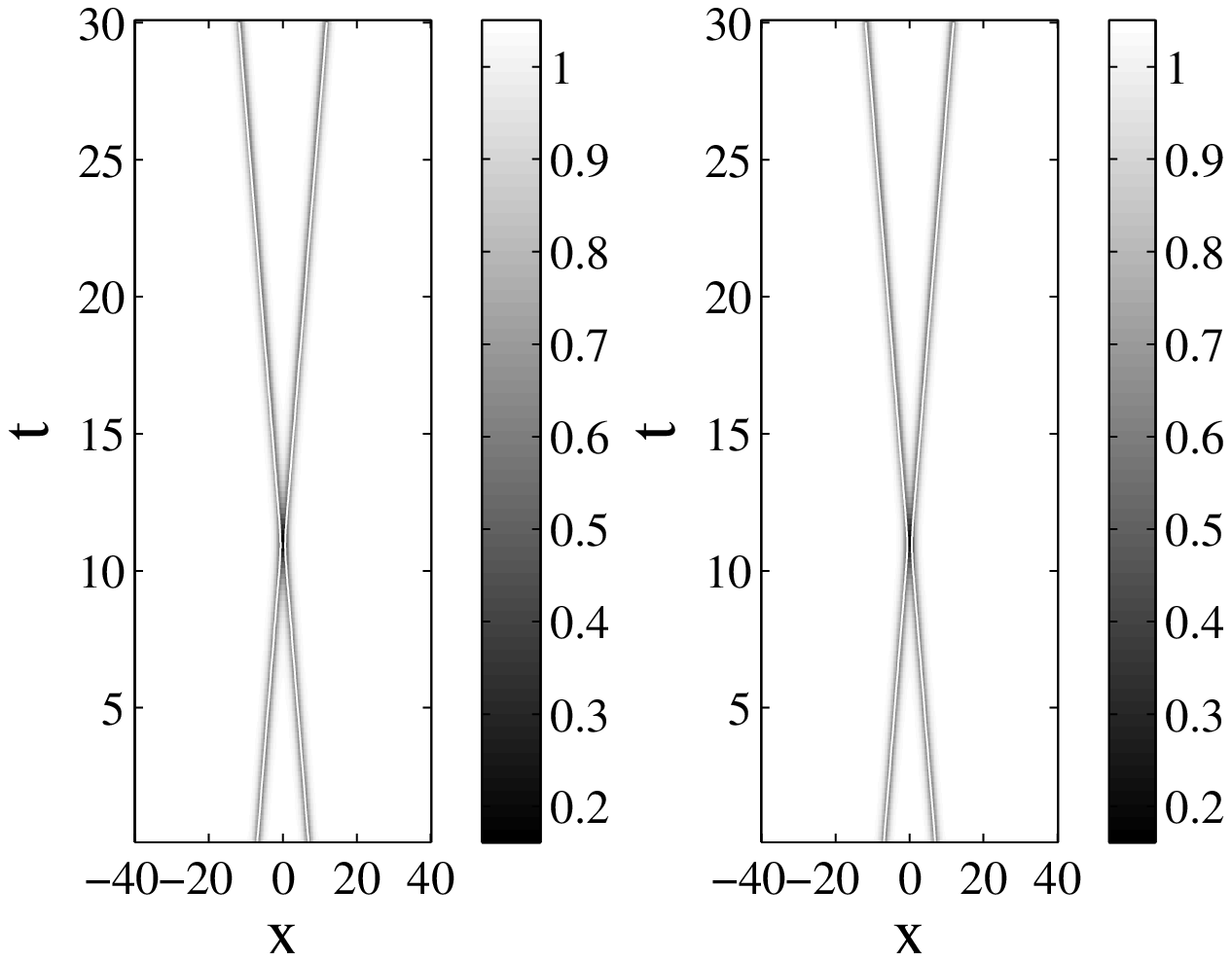}\label{Fig.64a}}
\subfigure[]{\includegraphics[width=8cm]{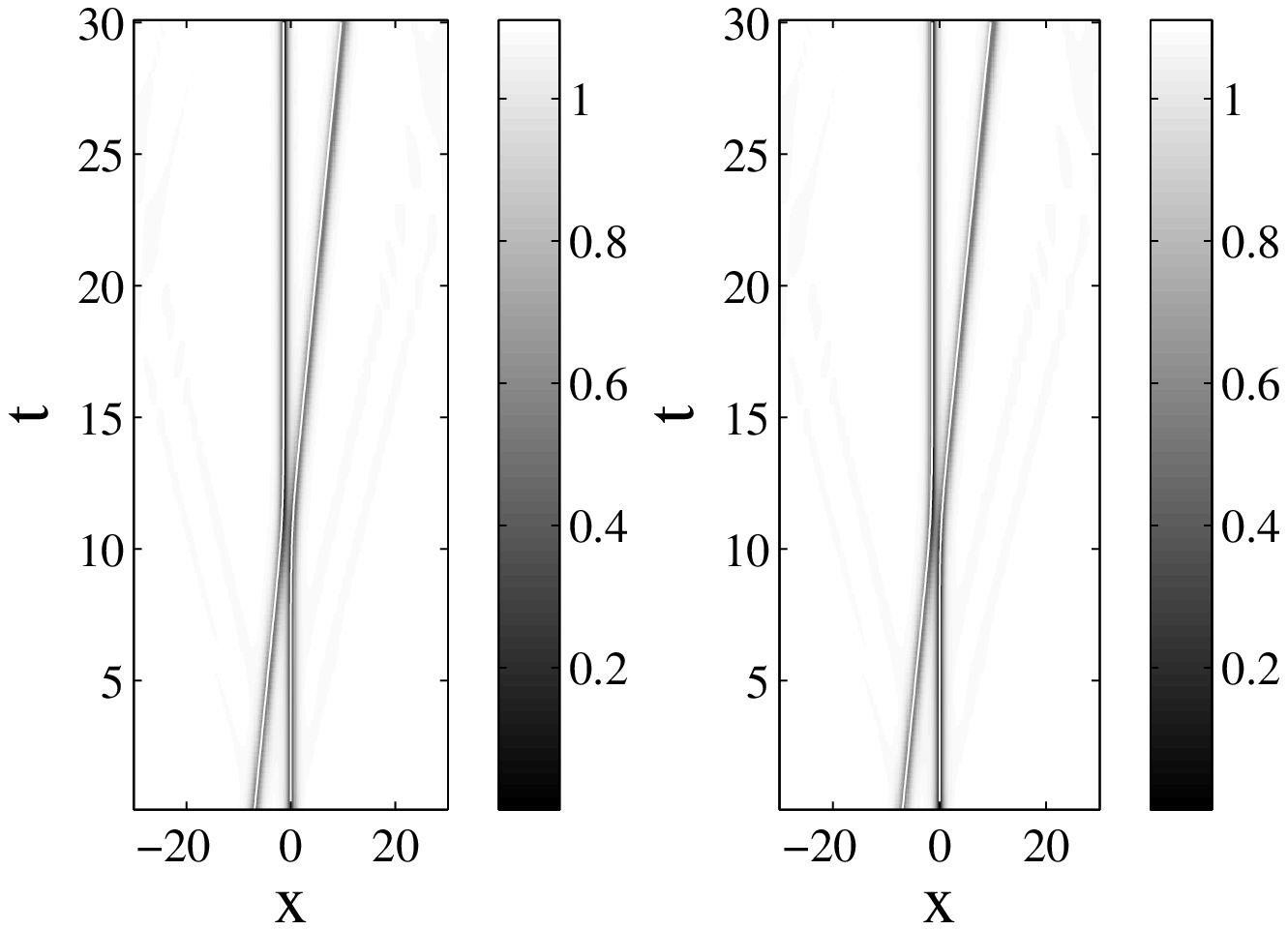}\label{Fig.66a}}
\caption{As Fig. \ref{Fig.59}, but for coupled dark solitons. In (a) the solitons are moving with velocities $v_1=-v_2=0.2$. The interparticle repulsion is dominant over the kinetic energies and the solitons are going away from each other after interaction. Due to an instability, they break down at approximately $t=120$. For (b) the coupled dark solitons are moving with velocities $v_1=-v_2=0.6$. The interparticle repulsion is suppressed by the kinetic energies of the solitons and they transmit through each other at the interacting point. In (c) one of the coupled solitons is static and the other is moving with velocity $v=0.5$. In all the cases $k=0.1$.}
\label{Fig.63}
\end{center}
\end{figure*}

Next, we consider the case $k\neq0$. Let us first consider the symmetric case in which a pair of coupled dark solitons interact with each other. The interaction scenarios for coupled dark solitons are almost similar to the scenarios for the uncoupled dark solitons in the previous section. The scenario for slow moving solitons is presented in Fig.\ \ref{Fig.63a}, where we can see that both coupled solitons are repelling each other and remain well separated for all time $t$. The only difference from the uncoupled case is that there is radiation emerging after interaction which was not seen in the uncoupled case. The reason for the emergence of this radiation is because the system is non-integrable. Also after a particular time the coupled dark solitons break up because they are unstable for the parameter values used for the interaction. Since a travelling uncoupled dark soliton is stable for all values of velocity, no radiation or break up could be seen. In the interaction scenario of the fast moving coupled solitons, they transmit through each other. Like the uncoupled case, their high kinetic energy overcomes the interparticle repulsion as shown in Fig.\ \ref{Fig.64a}. In an asymmetric interaction when one of the coupled dark solitons is static while the other is moving with some non-zero velocity $v$, the interaction scenario is similar to that of the uncoupled case and is shown in Fig.\ \ref{Fig.66a}. The trajectories of dips obtained by doing numerical simulations of Eq. (\ref{nEulerLageqs}) are compared with the trajectories found through direct numerical integration of Eq. (\ref{nls2}). The white solid curves in Fig.\ \ref{Fig.63} are the approximations obtained through Eq. (\ref{nEulerLageqs}) showing excellent agreement.

\subsection{Interaction of FAs in the absence of a magnetic trap}
\begin{figure}
\begin{center}
\includegraphics[width=8cm]{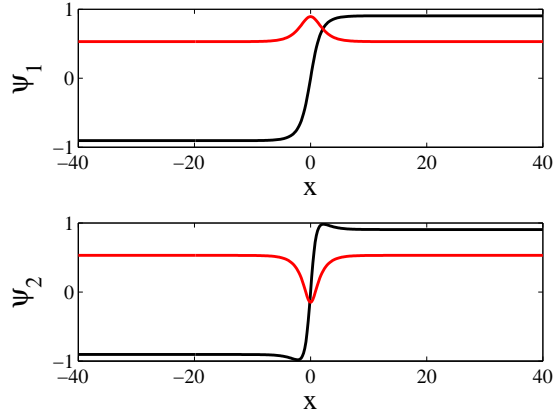}
\caption{Numerically obtained FAs travelling with velocity $v=0.2$ corresponding to $k=0.1$. The black curves represent the real parts while the red curves are the imaginary parts of $\psi_1$ and $\psi_2$, respectively.}
\label{Fig.67}
\end{center}
\end{figure}

\begin{figure*}[tbhp!]
\begin{center}
\subfigure[]{\includegraphics[width=8cm]{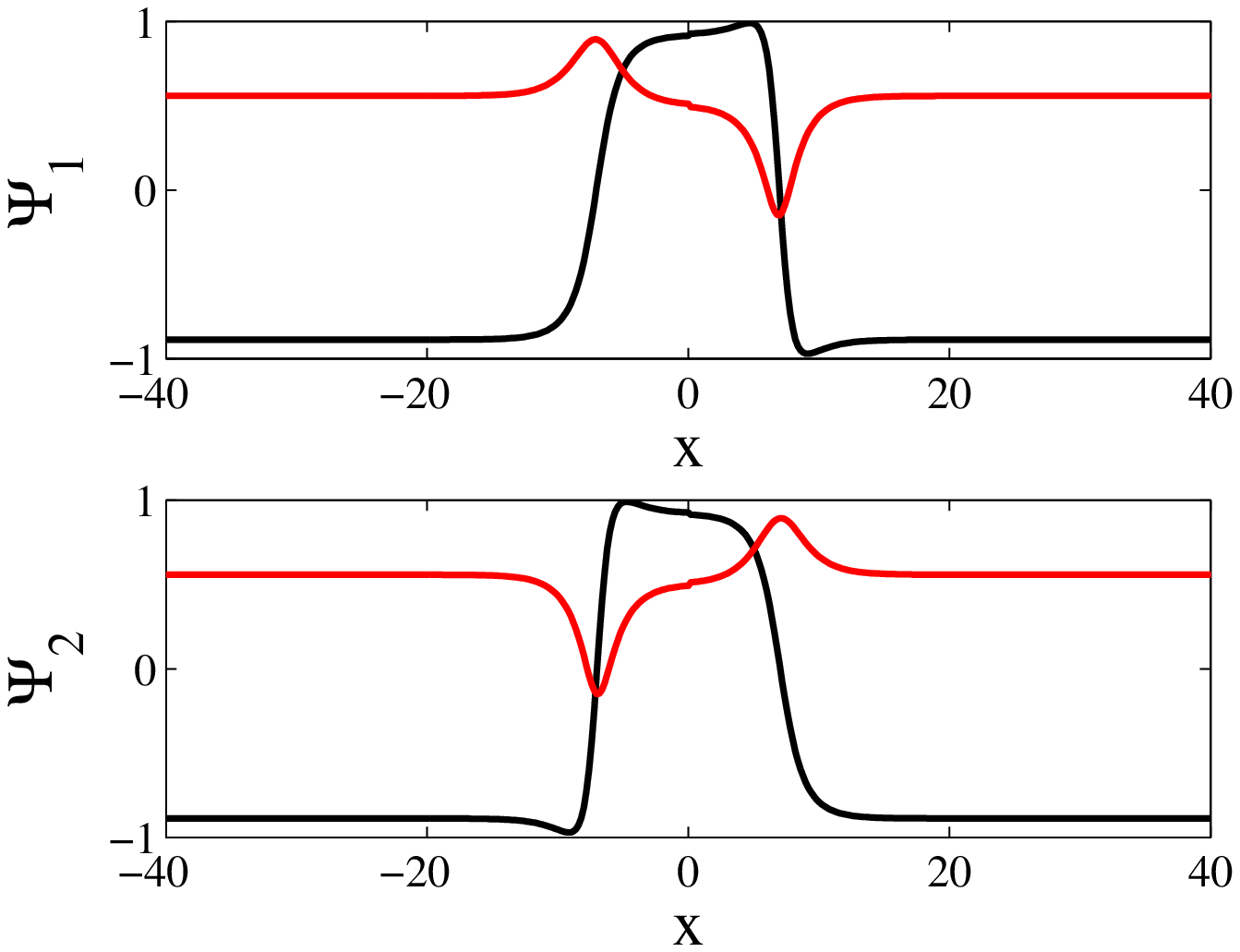}}
\subfigure[]{\includegraphics[width=8cm]{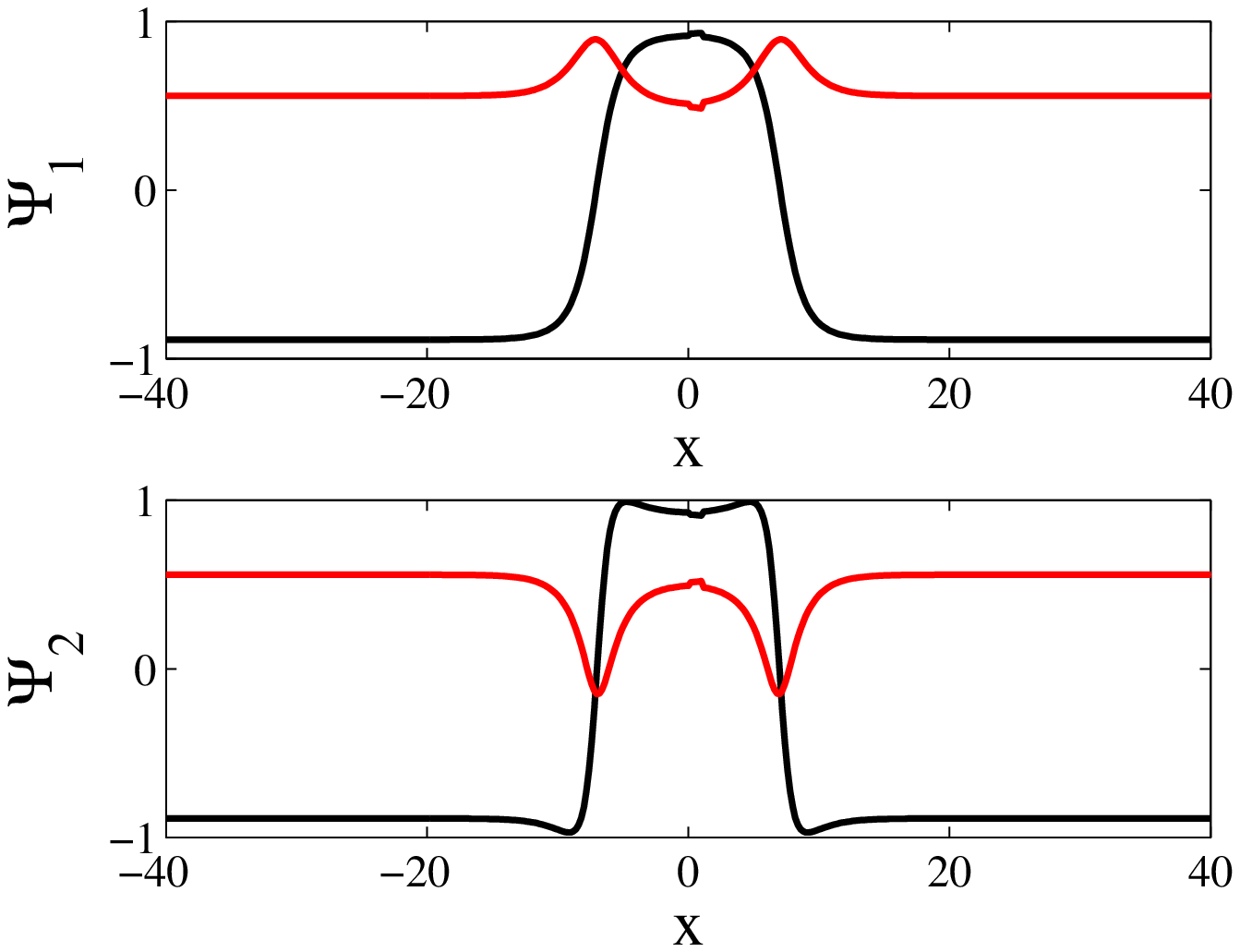}}
\caption{Profiles of an initial condition representing two coupled FAs travelling with velocities $v_1=-v_2=0.2$ corresponding to $k=0.1$. The figure in the upper panel represents the odd interaction, while figure in the lower panel is the even interaction.}
\label{Fig.68}
\end{center}
\end{figure*}

In this section, we will consider symmetric as well as asymmetric interactions of two FAs. An FA solution moving with velocity $v$ corresponding to a particular value of $k$ is shown in Fig.\ \ref{Fig.67}. We use a numerically obtained FA solution to construct a collision of two FAs. Since FAs are coupled solutions which consist of $\psi_1$ and $\psi_2$, there are two possibilities to connect two FA solutions. In the first possibility, $\psi_1$ and $\psi_2$ of the first FA are connected respectively to $\psi_1$ and $\psi_2$ of the second FA solution. In the second possibility, $\psi_1$ and $\psi_2$ of the first FA are connected respectively to $\psi_2$ and $\psi_1$ of the second FA solution. The combined pictures for both possibilities are shown in Fig.\ \ref{Fig.68}. From the symmetry of the imaginary parts, we refer to the first possibility as an odd symmetric interaction or ($+ -$)-configuration and the second possibility as an even symmetric interaction or ($+ +$)-configuration.

\begin{figure*}[tbhp!]
\begin{center}
\subfigure[]{\includegraphics[width=8cm]{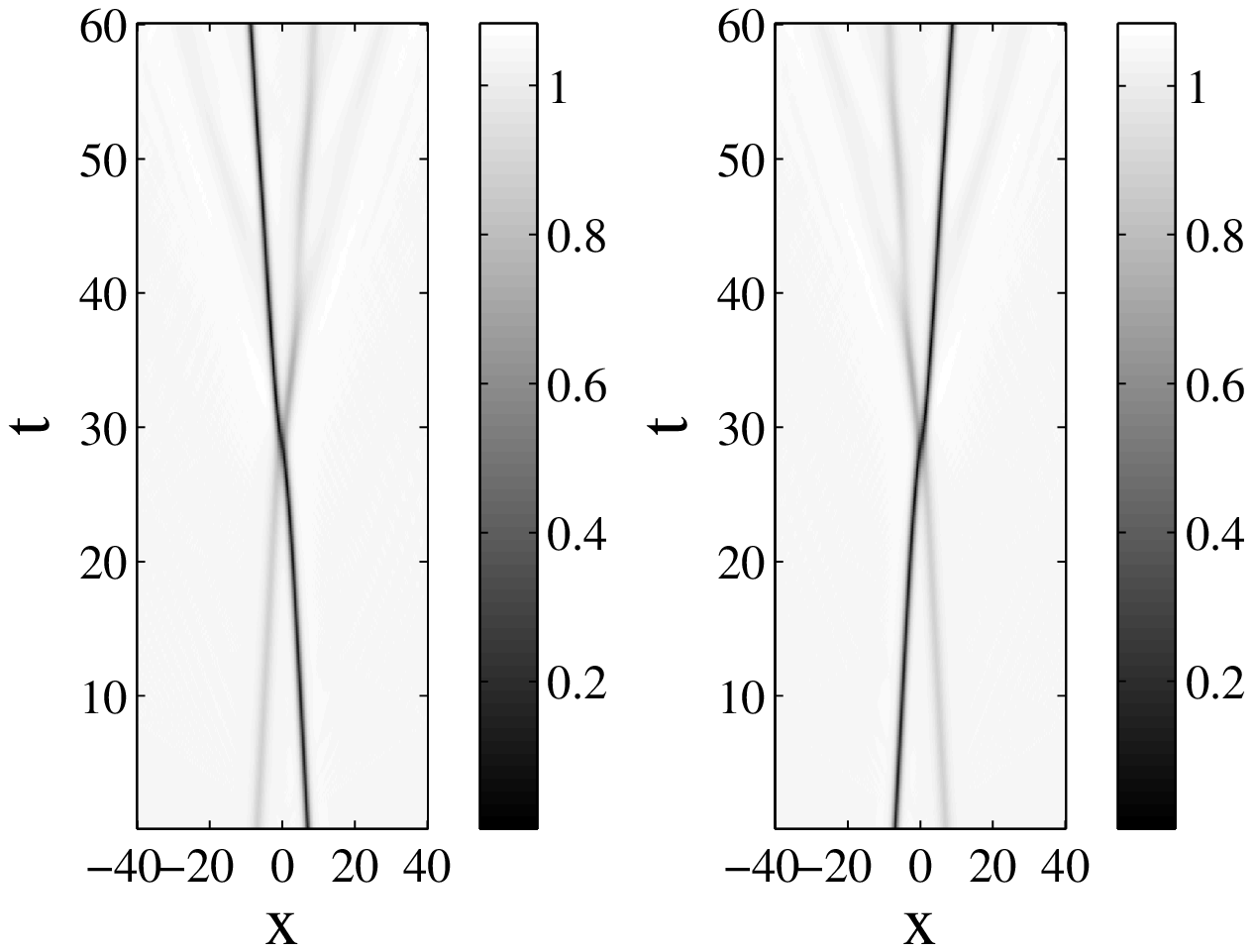}\label{Fig.69a}}
\subfigure[]{\includegraphics[width=8cm]{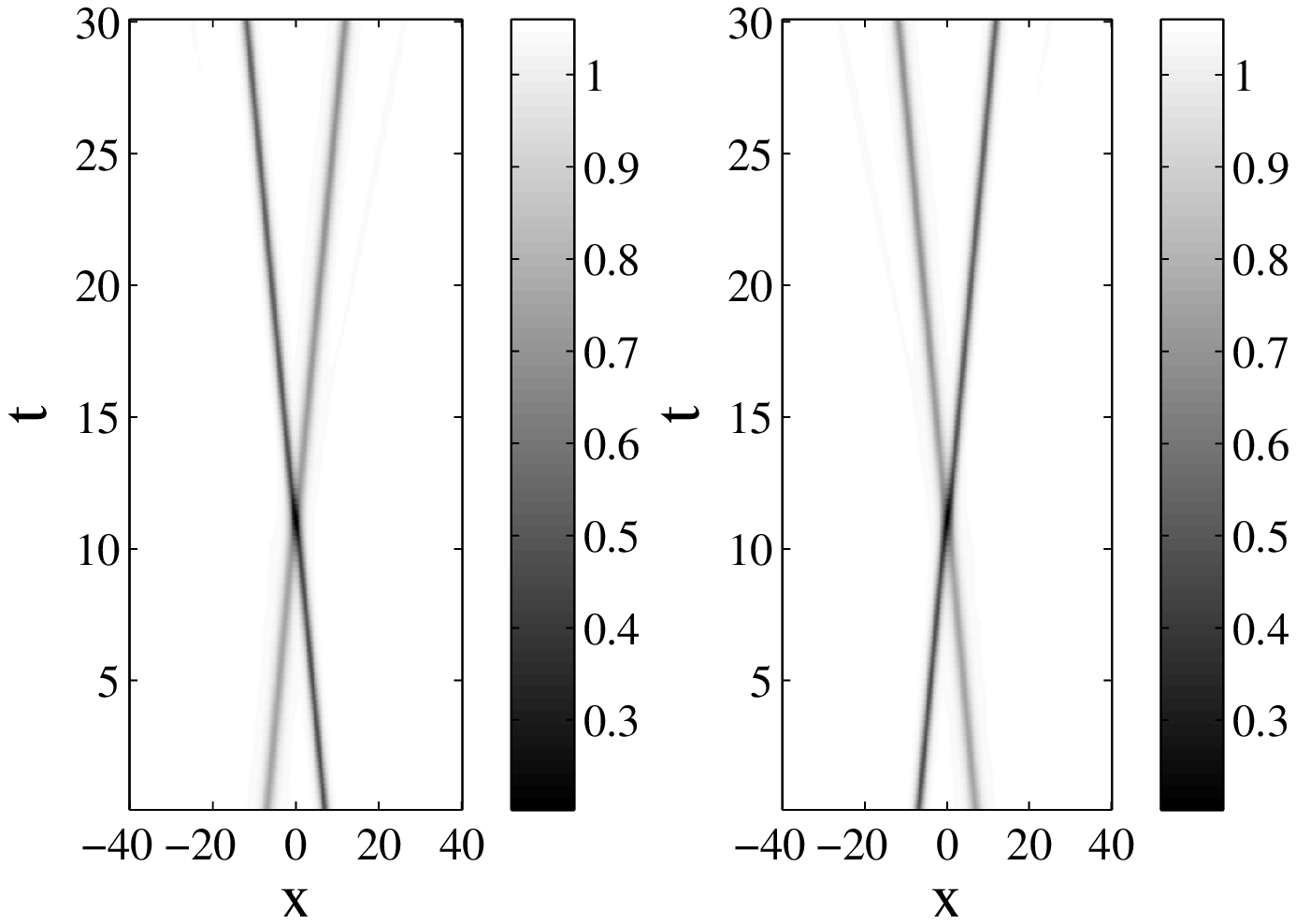}\label{Fig.70a}}
\caption{As Fig. (\ref{Fig.59}), but for the odd symmetric collision of FAs for (a) $v=0.2$ and (b) $v=0.6$. In both panels $k=0.1$.}
\label{Fig.69}
\end{center}
\end{figure*}

First, we discuss the odd symmetric interaction. In this case, two FA solutions initially localized at $x=\pm x_0$ move with opposite velocities. Shown in Fig.\ \ref{Fig.69a} is the interaction of relatively slow moving FAs. Both FAs can be characterized by an individual density minimum before and after collision while at the interacting point they exhibit a single nonzero density minima. Both FAs show attraction towards each other at the interacting point which results in the deflection of positions of dips of both solutions. Radiation emerges and phase shift is induced due to collision. Fig.\ \ref{Fig.70a} depicts an interaction of relatively high speed FA solutions. Neither of the FAs show any resistance during collision. Due to their high kinetic energies, both FAs pass through each other without shifting the phase and without showing any deflection in the trajectories of their dips. 

\begin{figure*}[tbhp!]
\begin{center}
\subfigure[]{\includegraphics[width=8cm]{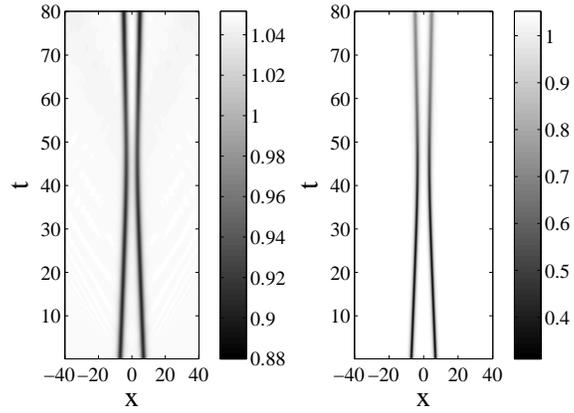}\label{Fig.71a}}
\subfigure[]{\includegraphics[width=8cm]{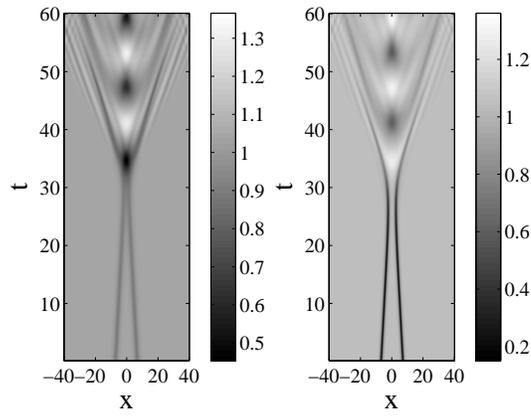}\label{Fig.72a}}
\subfigure[]{\includegraphics[width=8cm]{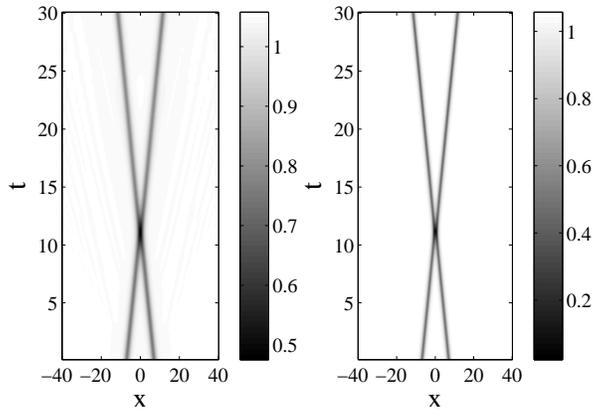}\label{Fig.74a}}
\caption{As Fig. (\ref{Fig.59}), but the even symmetric collision of FAs for (a) $v=0.1$, (b) $v=0.2$ and (c) $v=0.6$. In all panels $k=0.1$.}
\label{Fig.71}
\end{center}
\end{figure*}

Next, we consider the even symmetric interaction. An interaction of extremely slow moving FA solutions is displayed in Fig.\ \ref{Fig.71a}. The FAs come close to each other, but at the point of their closest proximity, they repel each other. Both the solutions can be identified by two individual density minima before and after the interaction as well as at the point of interaction. Another output from the interaction of two FAs moving with relatively slow velocities is shown in Fig.\ \ref{Fig.72a}. In this scenario, the solitons merge and form a breather similar to that reported in \cite{bara03} in a parametrically driven Schr\"odinger equation. Radiation emerging after the collision is clear. 
Shown in Fig.\ \ref{Fig.74a} is the collision of FA solution moving with relatively high velocities. The FA solutions collide with each other and become indistinguishable at the interacting point.

\begin{figure*}[tbhp!]
\begin{center}
\subfigure[]{\includegraphics[width=8cm]{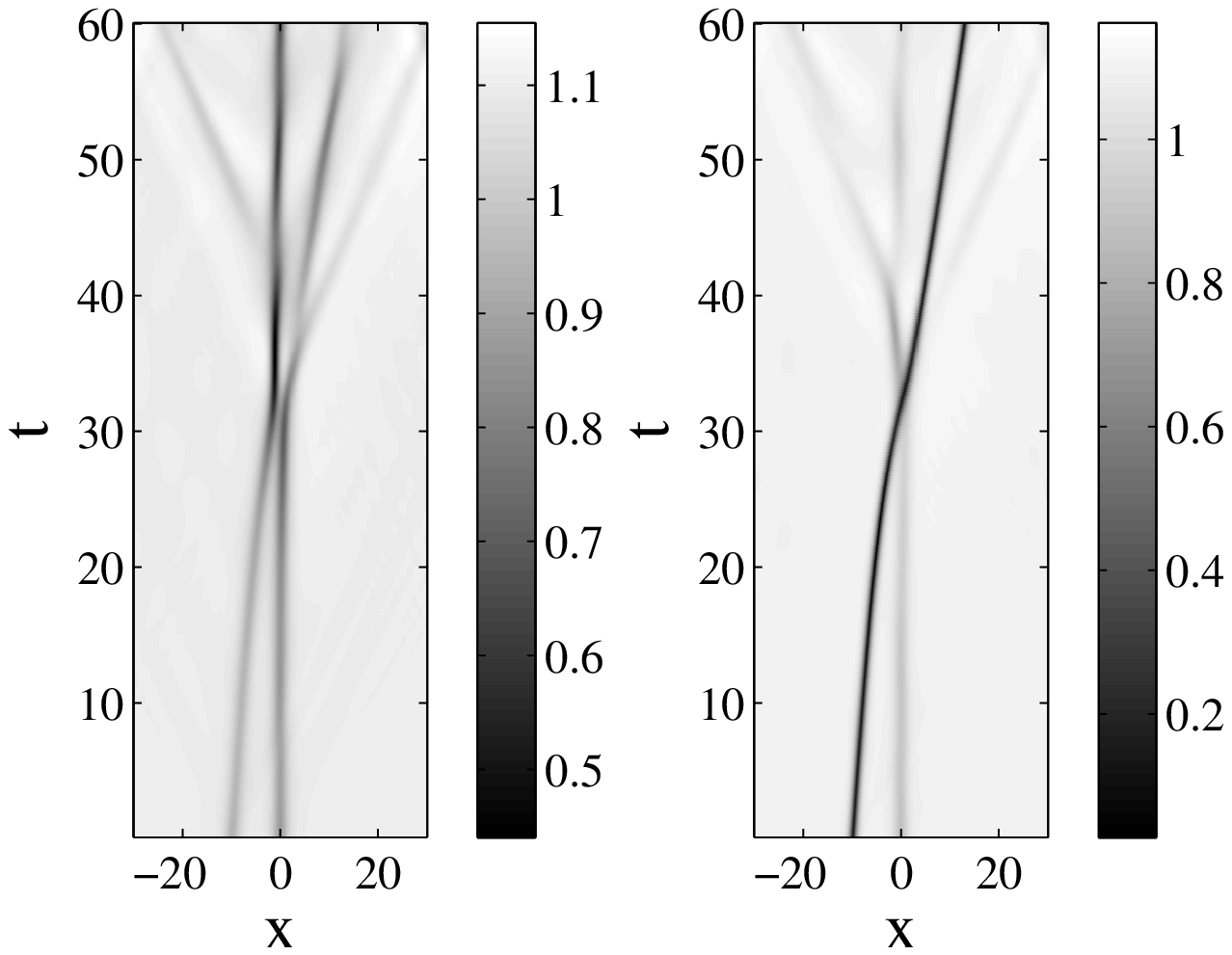}\label{Fig.76a}}
\subfigure[]{\includegraphics[width=8cm]{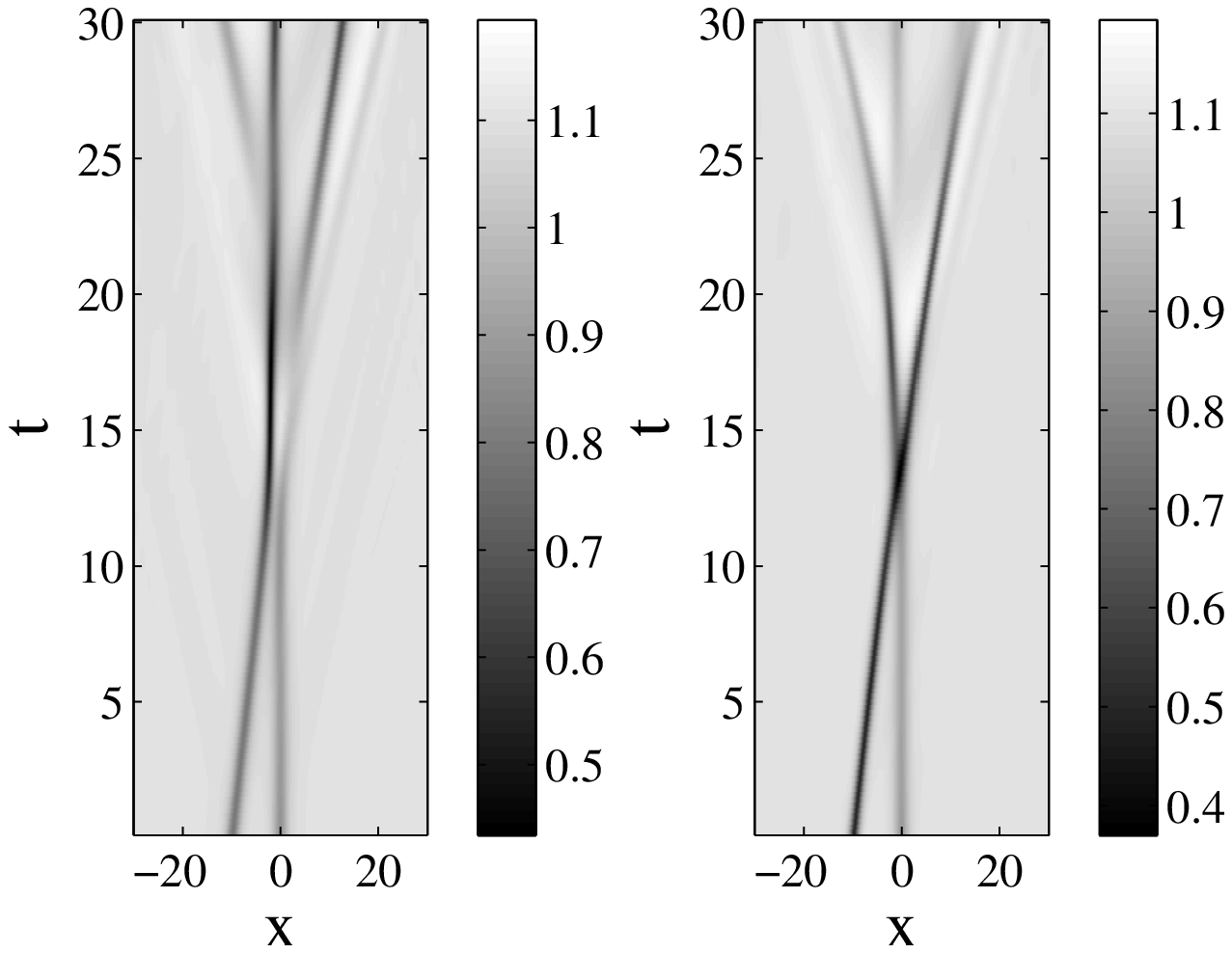}\label{Fig.77a}}
\caption{Numerical evolutions of the odd symmetric collisions of two FAs when one of them is static while the other is moving with velocity (a) $v=0.2$ and (b) $v=0.6$. In both panels $k=0.1$.}
\label{Fig.76}
\end{center}
\end{figure*}

\begin{figure*}[tbhp!]
\begin{center}
\subfigure[]{\includegraphics[width=8cm]{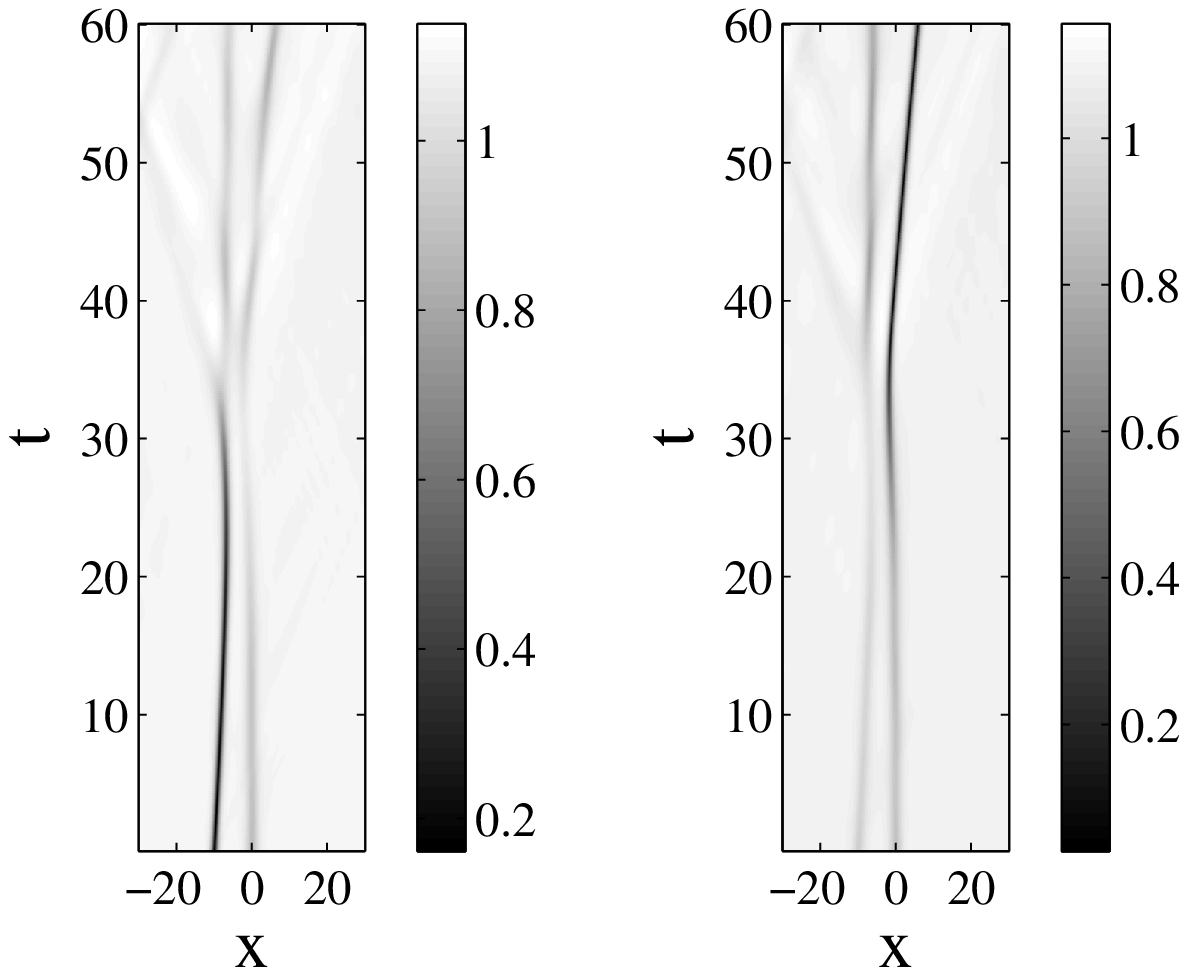}\label{Fig.78a}}
\subfigure[]{\includegraphics[width=8cm]{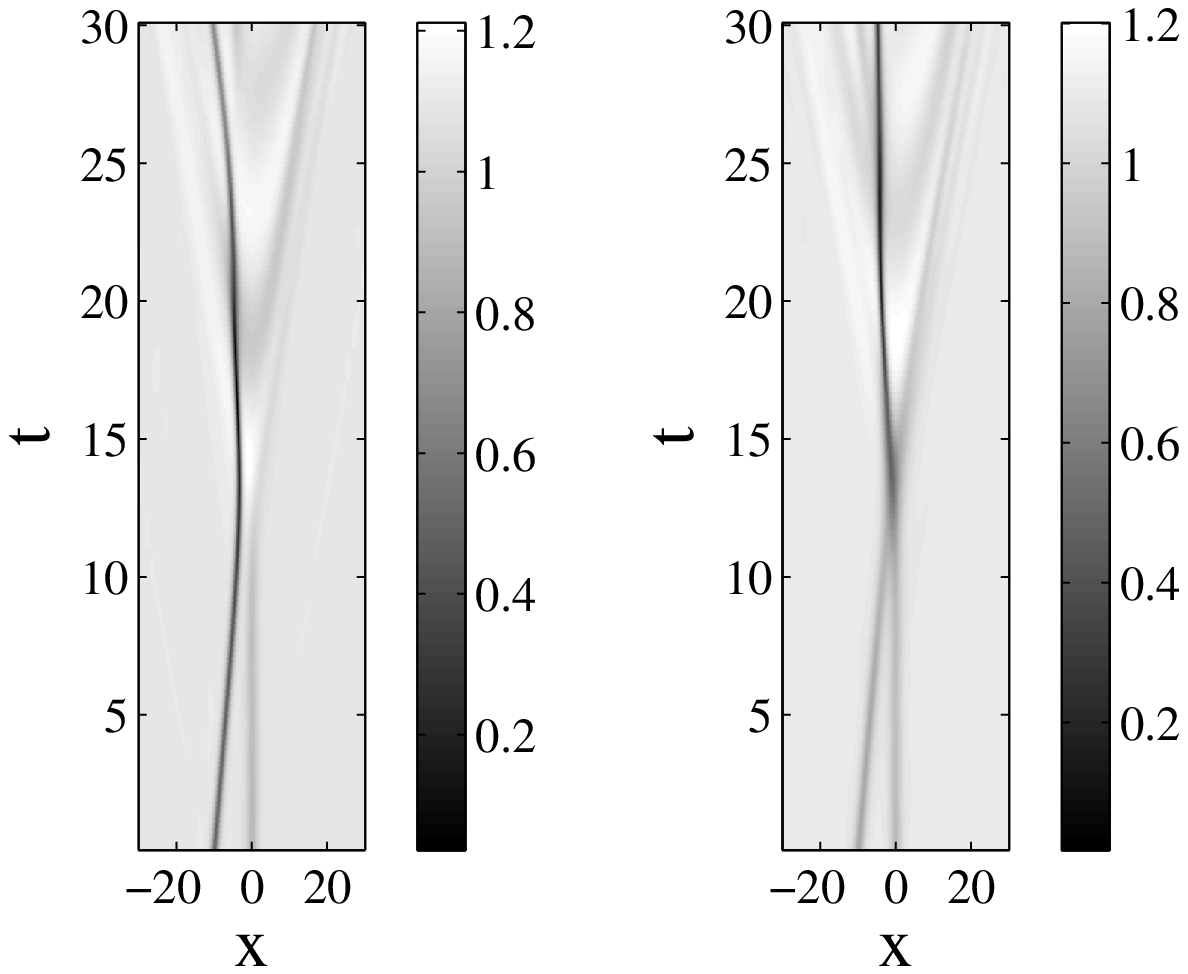}\label{Fig.79a}}
\caption{As Fig. (\ref{Fig.76}), but the even symmetric collision of FAs for (a) $v=0.2$ and (b) $v=0.6$. In both panels $k=0.1$.}
\label{Fig.78}
\end{center}
\end{figure*}

Finally, we consider asymmetric interactions of two FAs. We show in Fig.\ \ref{Fig.76a} an odd interaction of a relatively slow moving FA with a static FA. The travelling FA pushes the static FA away from the interacting point. This mean that the travelling FA transfers all
its kinetic energy to the static FA. Due to collision, radiation appears. In  Fig.\ \ref{Fig.77a} we show an interaction of a relatively fast moving FA solution with a static FA solution, where we obtain a similar behavior as before. Even interactions of a slow and a fast moving FA with a static FA are shown in Fig.\ \ref{Fig.78a} and Fig.\ \ref{Fig.79a}, respectively. From Figs.\ \ref{Fig.76} and \ref{Fig.78}, one can conclude that the collisions of a moving and a static FA are strongly inelastic. The radiation after the collision is so pronounce that it can be difficult to identify the outputs of the collisions.

\subsection{Stationary multiple FAs and dark solitons in the presence of magnetic trap}
In this section, we will consider the existence, stability and time-dynamics of multiple FAs and dark solitons in the coupled NLS equations (\ref{nls2}) in the presence of a magnetic trap. In particular, we consider stationary solutions of the governing equations.

To seek for static solutions in the time-independent framework of (\ref{nls2}), we use a Newton-Raphson continuation method. The spatial second order derivative is approximated using central finite differences with three-point or five-point stencils. At the computational boundaries, we use Neumann boundary conditions. In all the calculations, the grid spacing $\Delta x=0.2$ or smaller. Numerical linear stability analysis of a solution $\psi^{(0)}_j(x)$ is then performed by looking for perturbed solutions of the form
\[
\psi_j=\psi^{(0)}_j(x)+\epsilon[a_j(x)\,e^{i\lambda t}+b_j^*(x)\,e^{-i\lambda^* t}],\,j=1,2.
\]
Substituting the ansatz into the governing equation (\ref{nls2}) and keeping the linear terms in $\epsilon$, one will obtain a linear eigenvalue problem for the stability of $\psi^{(0)}_j$. The ensuing eigenvalue problem is then discretized using a similar finite difference scheme as above and solved numerically for the eigenfrequency $\lambda$ and corresponding eigenfunctions $a_j$ and $b_j$. It is then clear that $\psi^{(0)}_j(x)$ is a stable solution if the imaginary parts of all the eigenvalues vanish, i.e.\ $\textrm{Im}(\lambda)=0$.

\subsubsection{($+ -$)-configuration of FAs}
\begin{figure}
\begin{center}
\includegraphics[width=8cm]{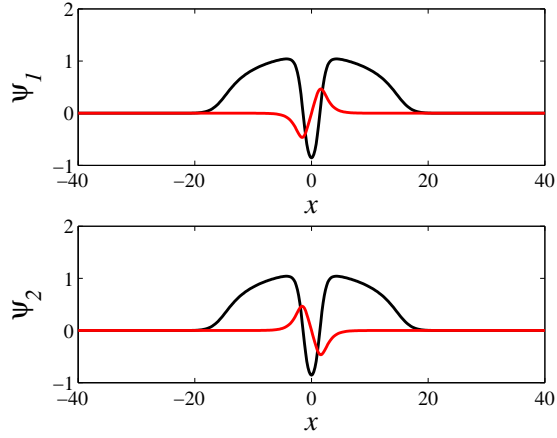}
\caption{Numerically obtained multiple FAs with a ($+ -$)-configuration for $\Omega=0.1$, $\rho_0=1$, $k=0.2$.}
\label{Fig.27}
\end{center}
\end{figure}

First, we consider the ($+ -$)-configuration of FA solutions of (\ref{nls2}), which is shown in Fig.\ \ref{Fig.27}. The amplitude of the imaginary parts of the solution reduces with $k$ and ultimately become zero at a critical value $k_{ce}$, when we obtain coupled dark solitons. The imaginary parts remain zero for all values of $k$ greater than or equal to $k_{ce}$.

\begin{figure}
\begin{center}
\includegraphics[width=8cm]{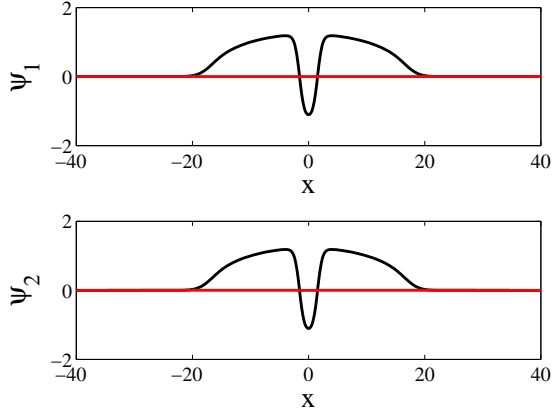}
\caption{Coupled dark soliton solutions for $\Omega=0.1$, $\rho_0=1$, $k=0.5$.}
\label{Fig.28}
\end{center}
\end{figure}

\begin{figure}
\begin{center}
\includegraphics[width=8cm]{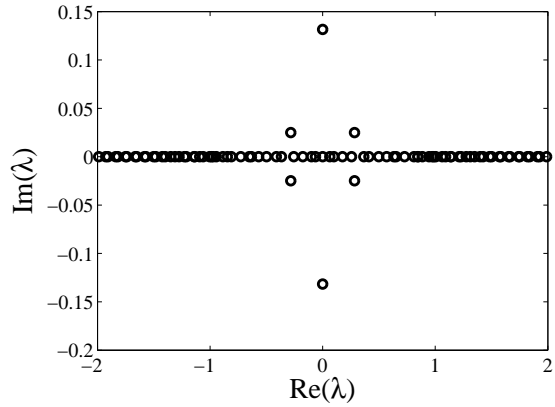}
\caption{The eigenvalue structure of the soliton in Fig. (\ref{Fig.27}) in the complex plane.}
\label{Fig.29}
\end{center}
\end{figure}

\begin{figure}
\begin{center}
\includegraphics[width=8cm]{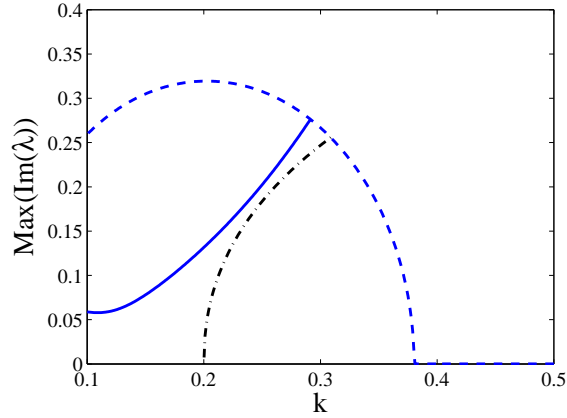}
\caption{The graph of $k$ and the maximum imaginary parts of eigenvalues for $\Omega=0.1$. The solid and dashed curves represent the trajectory of the most unstable eigenvalue for FAs and dark soliton as a function of $k$ respectively. The dash-dotted curve represents the approximation (\ref{omega2}) for the oscillation frequency of the ($+ -$)-configuration of FAs.}
\label{Fig.31}
\end{center}
\end{figure}

We have calculated the existence and stability of FA solutions for different non-zero values of trapping parameter $\Omega$. The critical value $k_{ce}$ for which this FA solution changes into dark soliton decreases with the increment of $\Omega$. The conversion of FAs into coupled dark solitons for $\Omega=0.1$ is shown in Fig.\ \ref{Fig.28}. The variation in the value of $\Omega$ also affects the stability of the solution. The critical coupling $k_{cs}$ where the solution becomes stable increases with $\Omega$. In this case, the value of $k_{cs}$ always remains greater than its corresponding value of $k_{ce}$. This shows that the ($+ -$)-configuration of FA solution is completely unstable for all values of $k$ where it exists. We note that the value of $k_{cs}$ corresponding to a specific value of $\Omega$ is actually the critical value for dark soliton at which it attains stability. The eigenvalue structure of FAs for a specific value of $\Omega$ is displayed in Fig.\ \ref{Fig.29} showing that most of the eigenvalues are real as they are lying on the horizontal axis, while few of them are complex. The most unstable eigenvalues are purely imaginary. The magnitude of the most unstable pair of eigenvalues increases for all $k\leq k_{ce}$ and then decreases with $k$ and ultimately becomes zero at $k=k_{cs}$. 
The stability curve is shown in Fig.\ \ref{Fig.31} by a solid curve. In the figure, we also present the stability curve of coupled dark solitons in dashed line. At $k=k_{ce}$ the solid and dashed curves meet. This corresponds to the situation when FAs turn into dark solitons, i.e. $k=k_{ce}$ is a pitchfork bifurcation point. The dark soliton becomes stable for $k \geq k_{cs}$. The dashed dotted curve shows the approximation (\ref{omega2}) obtained through a variational approach for $\Omega=0.1$, where one can see that a qualitatively good agreement is obtained.

\begin{figure}
\begin{center}
\includegraphics[width=8cm]{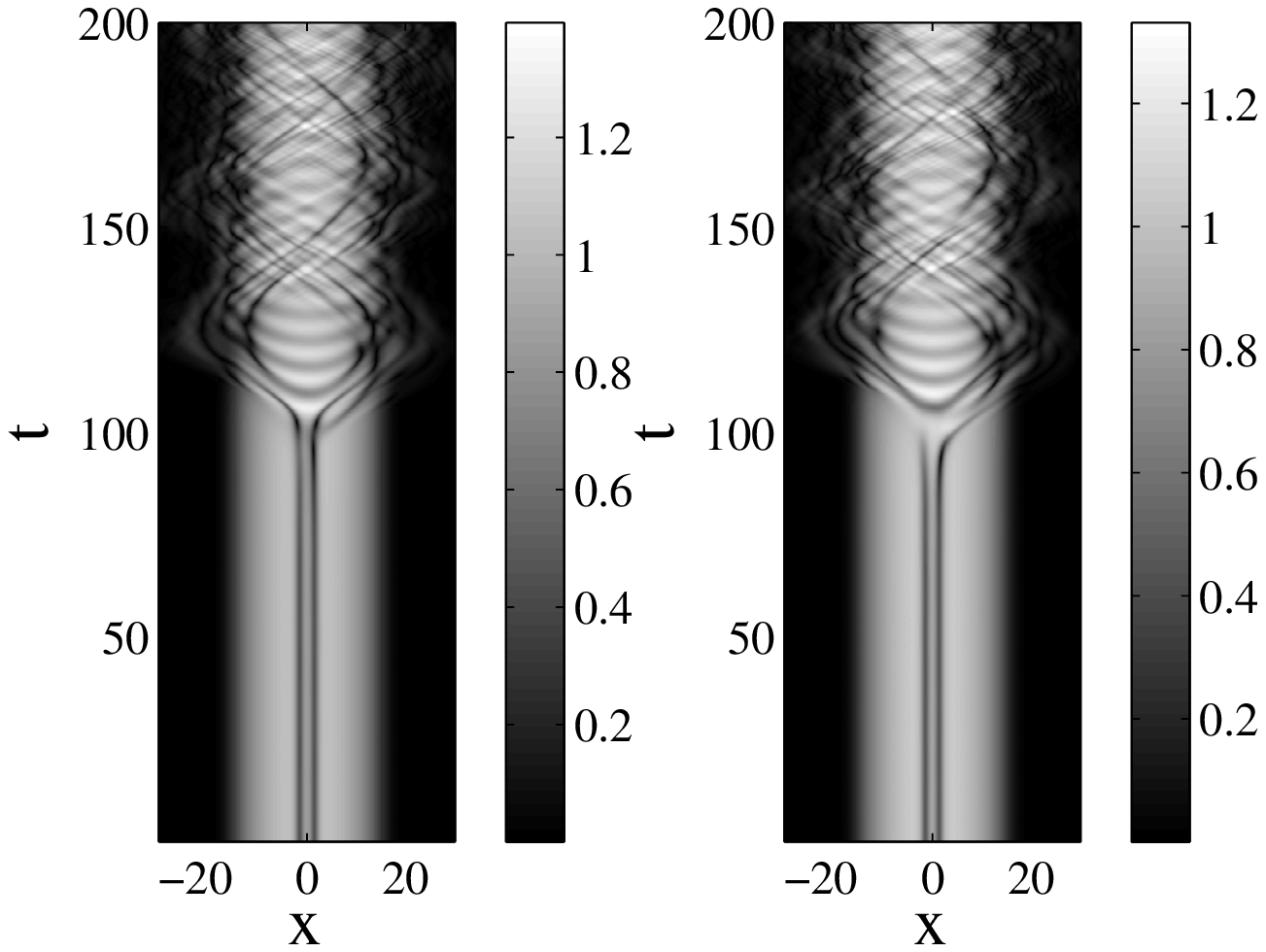}
\caption{Numerical evolution of the solution shown in Fig.\ \ref{Fig.27} for $\Omega=0.1$ and $k=0.2$.}
\label{Fig.32}
\end{center}
\end{figure}

In order to verify our results, we solve the time-dependent system (\ref{nls2}) numerically for the configuration of FA solutions above. For
$\Omega=0.1$ and $k=0.2$, the numerical evolution of the unstable FA solutions represented by $\psi_1$ and $\psi_2$ is shown in Fig.\ \ref{Fig.32}. Different from the collisions of two moving FAs with ($+ -$)-configuration that are attractive (see Fig.\ \ref{Fig.69}), the dynamics of unstable stationary FAs in here is rather repulsive. This can be seen in Fig.\ \ref{Fig.32} where at $t\approx100$, the FAs are moving from each other.

\subsubsection{($+ +$)-configuration of FAs}
\begin{figure}
\begin{center}
\includegraphics[width=8cm]{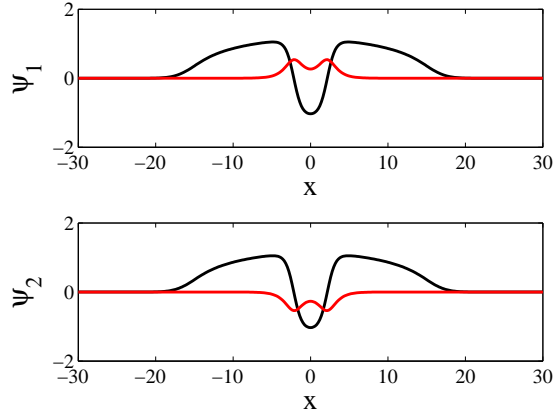}
\caption{Numerically obtained FAs for a ($+ +$)-configuration with $\Omega=0.1$, $\rho_0=1$, $k=0.25$.}
\label{Fig.35}
\end{center}
\end{figure}

\begin{figure}
\begin{center}
\includegraphics[width=8cm]{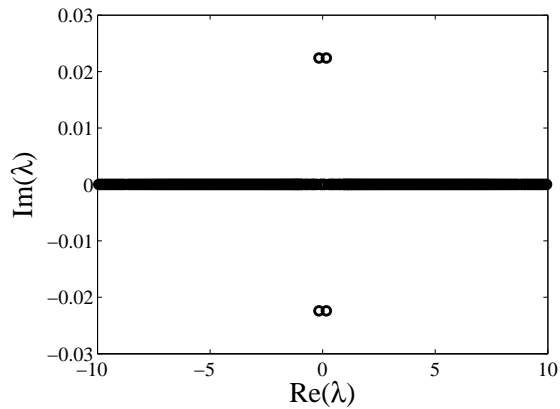}
\caption{The eigenvalue structure of the soliton in Fig.\ \ref{Fig.35}. All eigenvalues are real except two pairs of eigenvalues, which are complex, indicating the instability of the solution.}
\label{Fig.37}
\end{center}
\end{figure}

\begin{figure}
\begin{center}
\includegraphics[width=8cm]{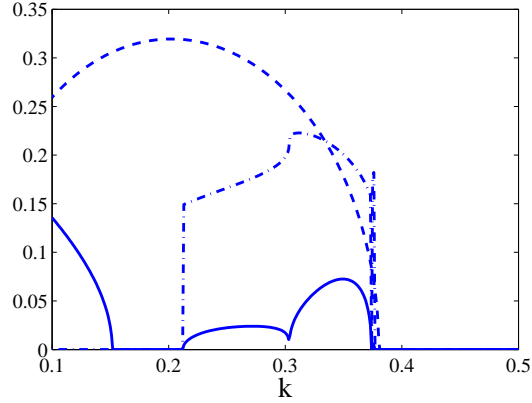}
\caption{The trajectory of the most unstable eigenvalue $\lambda_{max}$ corresponding to $\Omega=0.1$ for FAs solution with the ($+ +$)-configuration. The imaginary part of $\lambda_{max}$ is represented by the solid curve. The dashed line is the eigenvalue of coupled dark solitons (see Fig.\ \ref{Fig.31}). The dash-dotted curve shows the real part of $\lambda_{max}$ indicating an oscillatory instability when it is nonzero.}
\label{Fig.39}
\end{center}
\end{figure}

Finally, we consider FA solutions of (\ref{nls2}) with ($+ +$)-configuration as shown in Fig.\ \ref{Fig.35}. Similarly to the ($+ -$)-configuration, the imaginary parts of the solution reduces to zero with $k$ and at the critical value $k=k_{ce}$ coupled dark solitons are obtained. The critical value $k_{ce}$ increases with $\Omega$. The change in the value of $\Omega$ also changes the critical value $k_{cs}$ for the stability of this solution.  For $\Omega=0.1$, the ($+ +$)-configuration of FA solution is unstable for $k < 0.16$ due to two pairs of purely imaginary eigenvalues. The solution remains stable up to $k=0.21$, at which two pairs of unstable eigenvalues emerge from the spectrum. The imaginary part of the unstable eigenvalues becomes zero at $k=0.38$ showing that the solution is stable for $k \geq 0.38$. The eigenvalues structure for $\Omega=0.1$ and $k=0.25$ is shown in Fig.\ \ref{Fig.37}. Since this solution changes into a dark soliton at $k=0.38$, so the dark soliton is unstable for $k<0.38$ but becomes stable for $k\geq0.38$. The stability curve for $\Omega=0.1$ is shown in Fig.\ \ref{Fig.39} by solid line. The real part of the most unstable eigenvalue as a function of $k$ is displayed in the same figure by dashed line. At $k=k_{ce}$, the ($+ +$)-configuration in FA solutions merges with dark solitons, similarly to the case of ($+ -$)-configuration in a pitchfork bifurcation. Note that our analytical result (\ref{omega2}) cannot be used to approximate the instability of the ($+ +$)-configuration. It is because (\ref{omega2}) only yields purely imaginary eigenvalues while the instability of the solution here is oscillatory.

\begin{figure}
\begin{center}
\includegraphics[width=8cm]{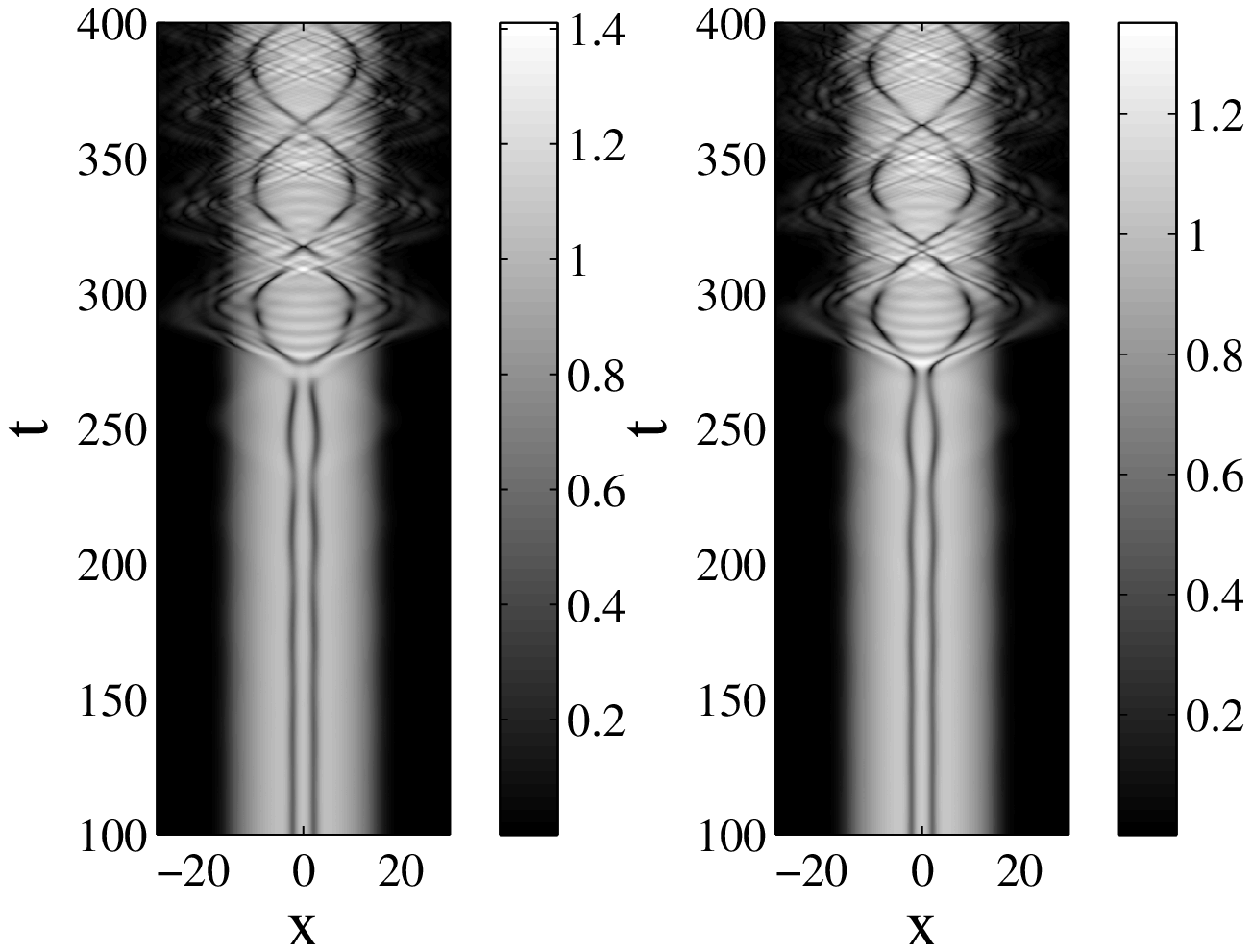}
\caption{Numerical evolution of the solution shown in Fig.\ \ref{Fig.35}.}
\label{Fig.40}
\end{center}
\end{figure}

The results obtained for the ($+ +$)-configuration of FA solutions above are also verified by direct numerical integration of the time-dependent system (\ref{nls2}). A typical evolution of unstable FA solutions is shown in Fig.\ \ref{Fig.40} for $\Omega=0.1$ and $k=0.1$. In a similar fashion as the ($+ -$)-configuration (see Fig.\ \ref{Fig.32}), the FAs repel each other. One difference between Fig.\ \ref{Fig.32} and Fig.\ \ref{Fig.40} is that in the latter case the break up is preceded by oscillations of the soliton pair. This is caused by the fact that the stability is oscillatory.

\section{Conclusion}
We have studied the existence and stability of multiple FAs and coupled dark solitons in linearly coupled Bose-Einstein condensates. In the absence of a harmonic trap, we have shown numerically that the interactions of the solitary waves are strongly inelastic, especially in the case of slow incoming velocities. Symmetric and asymmetric interactions of coupled dark solitons as well as FA solutions for different values of velocity were discussed. Interesting outcomes, such as breathers that do not exist in the uncoupled case, due to the inelastic collisions of FAs were observed. In the presence of a magnetic trap, bound states of solitons were shown to exist. The effects of variation of the trapping strength on the existence and stability of the multiple solitary waves were investigated numerically. It is found that for FAs with the ($+ -$)-configuration, the critical coupling for existence $k_{ce}$ decreases while the critical value for stability $k_{cs}$ increases with the magnetic strength $\Omega$. For the ($+ +$)-configuration, both $k_{ce}$ and $k_{cs}$ increase with $\Omega$. An analytical approximation was derived based on variational formulations to calculate the oscillation frequency of FA solutions with the ($+ -$)-configuration, where a qualitatively good agreement was obtained.

\newpage

\bigskip
\hrule
\smallskip
\hrule


\printindex
\end{document}